\documentclass[conference]{IEEEtran}

\usepackage{epsfig,endnotes}
\usepackage{varwidth}
\usepackage{algorithm}
\usepackage{algpseudocode}
\usepackage{balance}
\usepackage{xcolor}
\usepackage{nicefrac}
\usepackage{amsmath}
\usepackage{braket}
\usepackage{bm}
\usepackage{mathtools}
\usepackage{multirow}
\usepackage{bigdelim}
\usepackage{mathtools}
\usepackage{amssymb}
\usepackage{amsfonts}
\usepackage{algorithmicx}
\usepackage{indentfirst}
\usepackage{booktabs}

\usepackage{enumitem}
\usepackage{boxedminipage}
\usepackage{float}
\usepackage{graphicx}
\usepackage{caption}
\usepackage{subcaption}
\usepackage[normalem]{ulem}
\usepackage{xpatch}
\usepackage{xspace}
\usepackage{listings}
\usepackage{mathpartir}
\usepackage{makecell}
\usepackage[hyperfootnotes=false]{hyperref}%\usepackage{hyperref}
\usepackage{cleveref}
\usepackage{textcomp}
\usepackage{framed,mdframed}
\usepackage{tablefootnote}

\usepackage{flushend}

\usepackage{verbatim}
\usepackage{fancyvrb}
\usepackage{amsthm}

\usepackage{stmaryrd}

\usepackage{tablefootnote}

\newlength{\saveparindent}
\newlength{\saveparskip}
\newenvironment{tiretnospace}{%
\begin{list}{\hspace{2pt}\rule[0.5ex]{6pt}{1pt}\hfill}{\labelwidth=15pt%
\labelsep=5pt \leftmargin=20pt \topsep=3pt%
\setlength{\listparindent}{\saveparindent}%
\setlength{\parsep}{\saveparskip}%
\setlength{\itemsep}{0pt}}}{\end{list}}
 
\newenvironment{tiret}{%
\begin{list}{\hspace{2pt}\rule[0.5ex]{6pt}{1pt}\hfill}{\labelwidth=15pt%
\labelsep=5pt \leftmargin=20pt \topsep=3pt%
\setlength{\listparindent}{\saveparindent}%
\setlength{\parsep}{\saveparskip}%
\setlength{\itemsep}{0pt}}}{\end{list}}

\newcommand{\KwIn}[1]{\textbf{Input: }#1\\}
\newcommand{\KwOut}[1]{\textbf{Output: }#1\\}

\newcommand{\tensorflow}{{{TensorFlow}}\xspace}
\newcommand{\tool}{{\textsc{CrypTFlow}}\xspace}

\newcommand{\mpc}{{MPC}\xspace}

\newtheorem{theorem}{Theorem}

\newcommand{\namedref}[2]{\hyperref[#2]{#1~\ref*{#2}}\xspace}

\newcommand{\nc}[1]{{{\color{purple} nc: #1}}}

\definecolor{mypink}{rgb}{1,0.2,0.4}
\newcommand{\aseem}[1]{{\color{mypink}{[{\footnotesize {\bf Aseem:} { {#1}}}]}}}

\newcommand{\adv}{\mathcal{A}}

\newcommand{\simu}{\mathcal{S}}
\newcommand{\F}{\mathcal{F}}

\newcommand{\sign}{\mathsf{Sign}}

\newcommand{\fmpc}{\F_{\textrm{\tiny{mpc}}}}
\newcommand{\fattest}{\F_{\textrm{\tiny{attest}}}}

\newcommand{\gcommit}{\mathsf{Commit}}

\newcommand{\compute}{\mathsf{Compute}}
\newcommand{\ctr}{\mathsf{ctr}}

\newcommand{\st}{\mathsf{state}}
\newcommand{\transcript}{\mathsf{Transcript}}

\newcommand{\pk}{\mathsf{pk}}
\newcommand{\sk}{\mathsf{sk}}
\newcommand{\vk}{\mathsf{vk}}
\newcommand{\sksign}{\mathsf{sk}}
\newcommand{\protshtom}{\mathsf{Prot}_{\tiny{\mathsf{malicious}}}(P_1,\cdots,P_n)}
\newcommand{\verify}{\mathsf{Verify}}
\newcommand{\token}{\mathcal{T}}

\newcommand{\share}[3]{\langle #1\rangle^{#2}_{#3}}
\newcommand{\shareval}[2]{\langle #1\rangle^{#2}}

\newcommand{\reconst}[2]{\mathsf{Reconst}^{#1}(#2)}
\newcommand{\atrand}{\xleftarrow[\text{}]{\$}}
\newcommand{\bbZ}{\mathbb{Z}}
\newcommand{\bbF}{\mathbb{F}}
\newcommand{\relu}{\mathsf{ReLU}}
\newcommand{\shareconvert}{\mathsf{ShareConvert}}
\newcommand{\computemsb}{\mathsf{ComputeMSB}}
\newcommand{\prm}{p}

\newcommand{\maxpool}{\mathsf{MaxPool}}
\newcommand{\cryptflow}{\textsc{CrypTFlow}}
\newcommand{\resnet}{{\textsc{ResNet50}}\xspace}
\newcommand{\squeezenet}{\textsc{SqueezeNet}}
\newcommand{\densenet}{{\textsc{DenseNet121}}\xspace}

\newcommand{\etal}{{\em et al}.}

\usepackage{url}

\newenvironment{tffbox}{
\begin{figure}[ht!]\small
}{\end{figure}}

\usepackage{lipsum}

\newcommand\blfootnote[1]{%
  \begingroup
  \renewcommand\thefootnote{}\footnote{#1}%
  \addtocounter{footnote}{-1}%
  \endgroup
}
\ifCLASSINFOpdf
\else
\fi

\begin{document}
%
% paper title
% can use linebreaks \\ within to get better formatting as desired
\title{\cryptflow: Secure \tensorflow Inference} %Computation for Machine Learning Users}

% author names and affiliations
% use a multiple column layout for up to three different
% affiliations
\author{\IEEEauthorblockN{Nishant Kumar\IEEEauthorrefmark{1}}
\IEEEauthorblockA{Microsoft Research\\
t-niskum@microsoft.com}\\
\IEEEauthorblockN{Divya Gupta}
\IEEEauthorblockA{Microsoft Research\\
digup@microsoft.com}
\and
\IEEEauthorblockN{Mayank Rathee\IEEEauthorrefmark{1}}
\IEEEauthorblockA{Microsoft Research\\
t-may@microsoft.com}\\
\IEEEauthorblockN{Aseem Rastogi}
\IEEEauthorblockA{Microsoft Research\\
aseemr@microsoft.com}
\and
\IEEEauthorblockN{Nishanth Chandran}
\IEEEauthorblockA{Microsoft Research\\
nichandr@microsoft.com}\\
\IEEEauthorblockN{Rahul Sharma}
\IEEEauthorblockA{Microsoft Research\\
rahsha@microsoft.com}
}
\maketitle

\begin{abstract}\blfootnote{\IEEEauthorrefmark{1} Equal contribution}
We present \cryptflow, a first of its kind system that converts
\tensorflow inference code into Secure
Multi-party Computation (MPC) protocols at the push of a
button. To do this, we build three components. Our first component,
Athos, is an end-to-end compiler from \tensorflow
to a variety of semi-honest MPC protocols. The second
component, Porthos, is an improved semi-honest 3-party protocol
that provides significant speedups for \tensorflow like applications. Finally, to
provide malicious
secure MPC protocols, our third component, Aramis, is a novel
technique that uses hardware with integrity guarantees to convert any semi-honest
MPC protocol into an
MPC protocol that provides malicious security. 
The malicious security of the
protocols output by Aramis relies on integrity of the hardware and semi-honest security of \mpc.
Moreover, our system matches the inference accuracy of plaintext \tensorflow. 

We experimentally demonstrate
the power of our system by showing the secure inference of real-world
neural networks such as \resnet\ and \densenet\ over the
ImageNet dataset with running times of about 30 seconds for
semi-honest security and under two minutes for malicious
security. Prior work in the area of secure inference 
%(SecureML, MiniONN, ABY$^3$, SecureNN, CHET, Gazelle, and  Delphi)
 has been limited to
semi-honest security of small networks over tiny datasets such as MNIST or CIFAR. 
%In contrast, our largest network ({\textsc{ResNet200}}) has over 200 layers, 65 million parameters and over 1000 ImageNet classes.
Even on MNIST/CIFAR, \tool outperforms prior work.

\end{abstract}
%

% IEEEtran.cls defaults to using nonbold math in the Abstract.
% This preserves the distinction between vectors and scalars. However,
% if the conference you are submitting to favors bold math in the abstract,
% then you can use LaTeX's standard command \boldmath at the very start
% of the abstract to achieve this. Many IEEE journals/conferences frown on
% math in the abstract anyway.

% no keywords

% For peer review papers, you can put extra information on the cover
% page as needed:
% \ifCLASSOPTIONpeerreview
% \begin{center} \bfseries EDICS Category: 3-BBND \end{center}
% \fi
%
% For peerreview papers, this IEEEtran command inserts a page break and
% creates the second title. It will be ignored for other modes.
%%\IEEEpeerreviewmaketitle

\section{Introduction}
Secure multiparty computation (or MPC) allows a set of mutually distrusting parties to
compute a publicly known function on their secret inputs without revealing their
inputs to each other. This is done through the execution of a cryptographic protocol which guarantees that the protocol participants learn only the function output on their secret inputs and nothing else. 
\mpc has made rapid strides - from
being a theoretical concept three decades ago \cite{yao,gmw}, to now
being on the threshold of having real world impact.
One of the most compelling use cases for MPC is that of machine
learning (ML) - e.g. being able to execute inference over ML
algorithms securely when the model and the query are required to be
hidden from the participants in the protocol. There has been a flurry of
recent works aimed at running inference securely with MPC such as
SecureML~\cite{secureml}, MinioNN~\cite{minionn}, 
ABY$^3$~\cite{aby3}, CHET~\cite{chet},
SecureNN~\cite{securenn}, Gazelle~\cite{gazelle}, Delphi~\cite{delphi}, and so on.
 Unfortunately, these techniques are not easy-to-use by ML developers and have only been demonstrated on small deep
 neural networks (DNNs) on tiny datasets such as MNIST or CIFAR.
However, in order for MPC to be truly ubiquitous for secure inference
tasks, it must be both effortless to use and capable of handling large
ImageNet~\cite{imagenet} scale DNNs.

In this work, we present \cryptflow, a first of its kind system, that
converts \tensorflow \cite{tensorflow} inference code into \mpc protocols at the push of a button. By converting code in
standard \tensorflow, a ubiquitous ML framework that is used in production by
various technology companies, to \mpc protocols, we
significantly lower the entry barrier for ML practitioners and
programmers to use cryptographic \mpc protocols in real world
applications. We make the following four contributions:

\begin{tiret}

\item First, we provide a compiler, called {\em Athos}, from
  \tensorflow to a variety of secure computation protocols (both 2 and
  3 party) while preserving accuracy. In the absence of Athos, all prior works require {\em
    manually} re-implementing ML models in an MPC friendly low-level
  language/library, and hence, their evaluations have been limited to
 small benchmarks where this task is feasible.
%\dg{may be add sentence for ease to use} %demonstrating the secure inference of ImageNet scale neural networks.

\item Second, we provide a semi-honest secure 3-party computation protocol, {\em Porthos}, that outperforms all prior protocols for secure inference and enables us to execute, 
for the first time, the inference of ImageNet scale networks in {\em about 30 seconds}.
%\dg{quantify improvement.}

\item Third, assuming a minimally secure hardware  which guarantees
  the integrity of computations,  we show a novel technique, {\em
    Aramis}, that compiles any semi-honest secure MPC protocol to a malicious
  secure MPC protocol.
Aramis only relies on these integrity checks and assumes no confidentiality
guarantees for data residing within the hardware. 
% Prior works that combine \mpc and hardware are 
%secure only in weaker adversary models.
%The overhead of malicious security of 
Aramis
% based protocols is much
%lower compared to prior approaches, which 
enables the first
implementations of DNN inference secure against malicious
adversaries.
% Prior \mpc protocols are either much slower than Aramis
%or fail to provide security against malicious adversaries.

% with only 3-4x overhead.

\item Fourth, we demonstrate the ease-of-use, efficiency and scalability of \cryptflow\ by evaluating on  \\ (a) \resnet~\cite{resnet}, which won the ImageNet Large Scale Visual Recognition Challenge in 2015~\cite{imagenet}; \\ (b) \densenet~\cite{densenet}, a convolutional neural network that won the best paper at CVPR 2017.\\
% and has been used for real-world lung disease prediction on chest x-rays~\cite{lungdisease}
% and \\ (c) \squeezenet~\cite{squeezenet} with Fire modules.\\
\noindent 
These
 networks have heavily influenced the ML community with thousands of
 citations each. To demonstrate that \tool  is immediately useful in
 healthcare, we  also evaluate \tool on DNNs used for prediction of
 lung diseases and diabetic retinopathy.

\end{tiret}

Our toolchain and all of our benchmarks are publicly
available\footnote{\url{https://github.com/mpc-msri/EzPC}}.We now describe our results in more detail.

\subsection{Results}
\tool outperforms prior work on ease-of-use, scalability, and efficiency. It automatically compiles \tensorflow code to \mpc protocols with {\em no loss in classification accuracy}. This makes \cryptflow\ the first secure inference system to produce a Top 1 accuracy of $76.45\%$ and Top 5 accuracy of $93.23\%$ for predictions running securely on the ImageNet dataset. Furthermore, in the 3-party (3PC) setting, this can be done in about $30$ seconds with semi-honest security and about $2$ minutes with malicious security. 
Prior work in the area of secure inference has been limited to small networks over tiny datasets such as MNIST or CIFAR.
  Moreover, these implementations are limited to security against weaker semi-honest adversaries, that are assumed not to modify the code of the MPC protocol.
%The largest benchmark in the published literature on secure inference is Delphi's~\cite{delphi} 2-party (2PC) semi-honest secure %inference of 32 layer \textsc{ResNet}, with 0.46 million parameters, for 100 class CIFAR.  
In contrast, our largest network \textsc{ResNet-200} has 200 layers, 65 million parameters, over 1000 ImageNet classes, and the user can choose between semi-honest and malicious security -- the latter also protects against adversaries who can deviate from the MPC protocol specification arbitrarily. 
We have evaluated \tool on secure inference over DNNs that are at least an order of magnitude larger
than the state-of-the-art~\cite{delphi,chet,chameleon,securenn,secureml,gazelle,ezpc,minionn,aby3,nhe,xonn,quantizednn}.
Even on MNIST/CIFAR, \cryptflow\ has lower communication complexity and is more efficient than prior and concurrent works~\cite{securenn,aby3,chameleon,quantizednn}. 
Furthermore, \cryptflow\ is the first system to implement\footnote{ABY$^3$~\cite{aby3} provided a theoretical protocol to convert their semi-honest protocol into a malicious secure protocol on much smaller benchmarks than \cryptflow, but did not provide an implementation or experimental validation.} malicious security for secure DNN inference. 
We show that the overhead of Aramis over semi-honest protocols is small and varies between 25\% and 3X depending on the size of the computation.
%\rs{remove? For the case of small benchmarks such as $32-$bit addition using the GMW protocol, the overhead of Aramis is within $25\%$ of the semi-honest protocol.
%Prior ``crypto-only'' works on malicious secure MPC for such benchmarks had much higher overheads over their semi-honest secure counterparts~\cite{wrk17,wrk17b,krrw18}. 
%For large ImageNet scale benchmarks such as \resnet\ and \densenet, the overhead of Aramis is within YX of Porthos.}
Moreover, by very conservative estimates, Aramis based secure DNN inference is faster than state-of-the-art  malicious secure \mpc inference protocols~\cite{mpspdz} by at least an order of magnitude (and also the maliciously secure \mpc protocols for general computation~\cite{wrk17,wrk17b,krrw18}). Hence, on inference tasks, prior \mpc protocols are either much slower than Aramis  or fail to provide security against malicious adversaries. 

\subsection{Components of \cryptflow}
We describe the three components of \cryptflow\ next.
\\\\
\noindent\textbf{Athos (Section \ref{sec:athos}).} Athos is a compiler that compiles \tensorflow inference code to secure computation protocols. There are several challenges in doing so. For optimizations  (Section~\ref{sec:athosopt}), the compiler needs the dimensions of all the tensors occurring in the dynamic Python code.
The compiler is designed to be modular (Section~\ref{sec:athosmodularity}) and it provides facilities for plugging in various \mpc protocols.
%The output of Athos is a sequence of function calls where each function can be implemented by an appropriate \mpc protocol.
To demonstrate this modularity, we have implemented the following backends: ABY-based  2-party computation (2PC), Porthos-based semi-honest secure 3-party computation (3PC), and Aramis-based  malicious secure 3-party computation. 

The transformations implemented in Athos are sensitive to the performance of \mpc protocols. 
 For performance reasons all efficient secure computation protocols perform computation over fixed-point arithmetic - i.e., arithmetic over integers or arithmetic with fixed precision. This is in contrast to \tensorflow where computations are over floating-point values. Athos automatically converts \tensorflow code over floating-point values into code that computes the same function over fixed-point values. This compilation is done while {\em matching} the inference accuracy of floating-point code. 
%For this purpose, Athos leverages and builds upon the intermediary language SeeDot~\cite{seedot} while taking care that the conversion is done in a secure manner. While the SeeDot %compiler generates inherently insecure code, Athos is the first  secure compiler that automatically transforms \tensorflow floating-point code to fixed-point code for secure inference.
Prior works (\cite{secureml,minionn,gazelle,aby3,securenn,delphi}) in the area of running ML securely have performed this task by hand with significant losses in accuracy over floating-point code.
% For example, it is trivial~\cite{tftutorial} to obtain a floating-point DNN with over $99\%$ accuracy on classifying handwritten digits as $0,1,\cdots,9$. However, SecureML~\cite{secureml} %works with a hand constructed fixed-point DNN which has only $94\%$ accuracy to classify digits as 0 or 1.
 Although these fixed-point conversions are feasible to do manually for one or two small benchmarks, this task quickly becomes intractable for large benchmarks and needs to be repeated for every new benchmark. Athos automates this tedious and error prone task.
% Athos works by ``sweeping through'' various precision levels to estimate the best precision.
% By doing so, Athos exhibits negligible loss in accuracy over its insecure counterparts. In some instances, Athos even improves the inference accuracy! 
%This design of Athos addresses the challenge of modularity and makes it easy
%to incorporate new \mpc protocols ) and compiler optimizations.
\\\\
\noindent\textbf{Porthos (Section \ref{sec:porthos}).} 
Porthos is an improved semi-honest 3-party secure computation protocol (tolerating one corruption) that builds upon SecureNN~\cite{securenn}. 
Porthos makes two crucial modifications to SecureNN. 
First, SecureNN reduces convolutions  to matrix multiplications and  invokes the Beaver triples~\cite{beaver} based matrix multiplication protocol. 
When performing a convolution with filter size $f\times f$ on a matrix of size $m\times m$, the communication is roughly $2q^2f^2+2f^2+q^2$ elements in the ring $\bbZ_{2^{64}}$, where $q = m-f+1$. 
Porthos computes these Beaver triples by appropriately reshaping $m\times m$ and $f\times f$ matrices. 
This reduces the communication to roughly $2m^2+2f^2+q^2$ ring elements. 
Typically the filter size, $f$, is between 1 and 11 and the communication of Porthos can be up to two orders of magnitudes lower than SecureNN. 
%Let $P_0, P_1$ and $P_2$ denote the 3 parties in the protocol.
Additionally, in SecureNN, the protocols for non-linear layers (such as Rectified Linear Units (ReLU) and MaxPool)  require the third party to send secret shares to the first two parties. 
In Porthos, we cut this communication to half by eliminating the communication of one of these shares. 
This reduces the communication in the overall ReLU and MaxPool protocols by 25\%.
Thus, by reducing the communication in both linear convolution layers and non-linear layers, the communication in Porthos is several  GBs lower than SecureNN  (Table~\ref{tab:porthosvssecurenn}). 
%we make the observation that communication of one of these shares can be eliminated (as it is purely random) and can be pre-shared between the two parties using a PRF key. Additionally, we also ``load balance'' this pre-sharing across the two pairs of parties ($P_2,P_1$ and $P_2,P_0$) as these functions are always computed in large batches in machine learning applications. 
%Finally, Porthos optimizes local computations of parties to derive maximum benefit from these communication reductions.
\\\\ 
\noindent\textbf{Aramis (Section \ref{sec:aramis}).}
%\dg{remove definitions from here?}
%Semi-honest secure MPC protocols assume that the protocol participants
%follow the protocol specification honestly and compute every message
%of the protocol correctly with respect to their input and the protocol
%history. On the other hand, maliciously secure MPC protocols make no
%such assumptions on the adversary and are guaranteed to be secure even
%when protocol participants deviate arbitrarily from the
%protocol. 
Obtaining maliciously secure MPC protocols through
cryptography can often be challenging and expensive -- typically some
sort of ``proof of honest computation'' must be provided by the
parties for every step of the protocol. We present a novel technique, called Aramis, that compiles \mpc
protocols secure against semi-honest adversaries into \mpc protocols
that are secure against malicious adversaries, by leveraging secure
hardware. 
We only require the hardware to provide
code attestation and a secure signing functionality (that we use to
sign and verify the protocol messages). Aramis has two attractive features: (a) it works in a strong adversarial threat model; and (b) it serves as a general technique that can work on a variety of semi-honest secure MPC protocols. In more detail:

\begin{tiretnospace}
\item[(a)] The threat model of Aramis is significantly stronger than the prior
work on MPC using secure
hardware~\cite{vc3, obliviousmpml, GuptaFC16, BahmaniFC17,gcsgx,ndss1,ndss2,ndss3,slalom,opaque,chiron}.
Specifically, in our threat model, not only is the host operating system outside the Trusted Computing Base, but it is also allowed to observe
the entire state of the hardware (including user data). 
In contrast, for security of the protocol, the prior works require that the hardware hides the state from  the host and even if data is decrypted and computed upon inside the hardware, it cannot be viewed by the host. In Section~\ref{sec:aramis}, we describe the Aramis threat model in
more detail, formalize the secure hardware as an ideal functionality,
provide a formal description of the malicious secure MPC protocols,
and formally prove their security. The ideal functionality can potentially
 be realized using various hardware platforms that provide
code attestation and signing, e.g., STM32H7, MediaTek MT3620, CEC1702, ARMTrustZone, Intel's SGX, etc.
We provide a proof-of-concept 
implementation of Aramis by using SGX as the underlying secure
hardware. 
\item[(b)] Aramis is general and can be applied to any semi-honest secure \mpc protocol. To demonstrate this, we
derive malicious secure MPC protocols from both semi-honest GMW (2
party protocol)~\cite{gmw} and
Porthos (3 party protocol). Porthos compiled with Aramis gives the
first experimentally vetted maliciously secure protocol for neural
network inference with at most 3X overhead over semi-honest security. 
While these were the semi-honest protocols we applied Aramis to, one could potentially obtain performant maliciously secure variants of several other recent semi-honest secure inference protocols (e.g. \cite{gazelle, delphi, nitin}), and \mpc protocols for other applications~\cite{krtwpsi, pisgoogle}.
\end{tiretnospace}

\subsection{Organization of the paper} We provide an end-to-end walkthrough of our system to illustrate the overall toolchain in Section \ref{sec:toolchain}. %After discussing preliminaries related to neural networks, security, and SGX in Section \ref{subsec:preliminaries}
In Section \ref{sec:athos}, we describe our compiler Athos. Section \ref{sec:porthos} describes our improved 3-party semi-honest secure protocol for neural networks. We describe Aramis that compiles any semi-honest secure protocol into a malicious secure protocol, in Section \ref{sec:aramis}. We present all our experimental results in Section \ref{sec:experiments}, related works in Section \ref{sec:related} and conclude in Section \ref{sec:conclusion}.

\section{Motivating Example}\label{sec:toolchain}

\begin{figure}
  \includegraphics[width=\linewidth]{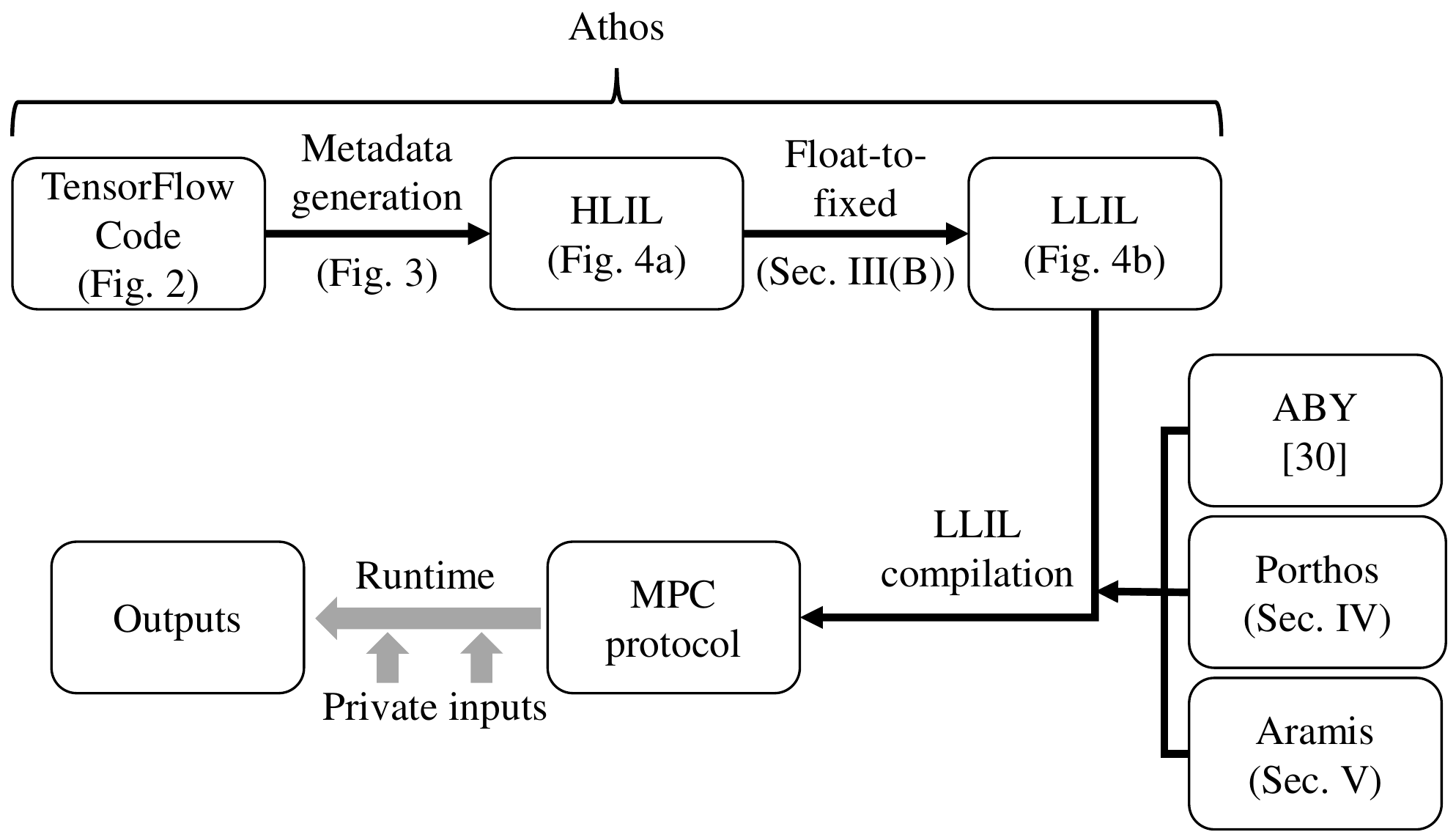}
  \caption{\cryptflow: End-to-end toolchain}
  \label{fig:cryptflowtoolchain}
\end{figure}

In this section, we describe the end-to-end working of \cryptflow\ through an example of logistic regression. The high-level toolchain is shown in Figure \ref{fig:cryptflowtoolchain}. We describe how code compilation happens from \tensorflow to \mpc protocols. 

%% The developer would write code in \tensorflow, which is the first step
%% of the toolchain.

\begin{figure}
  \centering
  \resizebox{0.49\columnwidth}{!}{
    \begin{subfigure}{0.4\columnwidth}
      \centering
      \includegraphics[scale=0.8]{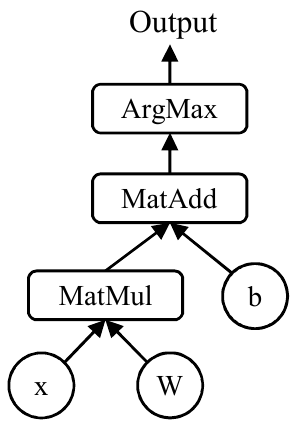}
      \caption{}
      \label{fig:tfGraphDef}
    \end{subfigure}
  }
  \resizebox{0.38\columnwidth}{!}{
    \begin{subfigure}{0.4\columnwidth}
      \centering
      \def\arraystretch{1.2}
      \begin{tabular}{|p{1.2cm}|p{1.6cm}|}
        \hline
        Node & Outgoing \\ & dimensions \\ \hline \hline
        x & $1 \times 784$\\ \hline
        W & $784 \times 10$\\ \hline
        MatMul & $1 \times 10$ \\\hline
        b & $1 \times 10$ \\\hline
        MatAdd & $1 \times 10$ \\\hline
        ArgMax & $1 \times 1$ \\\hline
      \end{tabular}
    \caption{}
    \label{fig:tfGraphMetadata}
    \end{subfigure}
  }
  \caption{Logistic Regression: (a) \tensorflow graph definition (b) Metadata consisting of graph nodes and their outgoing dimensions}
  \label{fig:lrtfgraphdump}
\end{figure}

\begin{figure}
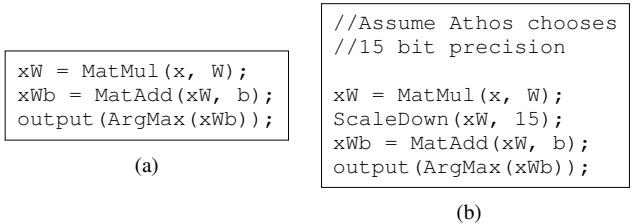

  \centering
  \resizebox{0.47\columnwidth}{!}{
    \begin{subfigure}{0.49\columnwidth}
      \centering
      \setlength{\fboxsep}{1.7mm}
      \fbox{\BUseVerbatim[fontsize=\small]{HLIL_LR_Verbatim}}
      \caption{}
      \label{fig:lrseedot}
    \end{subfigure}
  }
  \resizebox{0.47\columnwidth}{!}{
    \begin{subfigure}{0.49\columnwidth}
      \centering
      \setlength{\fboxsep}{1.6mm}
      \fbox{\BUseVerbatim[fontsize=\small]{LLIL_LR_Verbatim}}
      \caption{}
      \label{fig:lrezpc}
    \end{subfigure}
  }
  \caption{Logistic Regression in (a) floating-point: HLIL syntax (b) fixed-point: LLIL syntax}
\end{figure}

The \cryptflow\ toolchain takes as input code written in vanilla
\tensorflow. For example, consider the code snippet for
logistic regression over MNIST
dataset in \tensorflow as shown in Figure \ref{fig:lrtf}. 
Our compiler
% compiles this code to \mpc protocols using the following
%sequence of steps. It
 first generates the
\tensorflow graph dump (as shown in Figure \ref{fig:tfGraphDef}) as
well as metadata to help compute the dimensions of all the tensors
(Figure \ref{fig:tfGraphMetadata}). \ref{subsec:athosfrontend}
provides more details on the frontend. Next, the \tensorflow graph
dump is compiled into a high-level intermediate language HLIL. The
code snippet for logistic regression in HLIL is shown in Figure
\ref{fig:lrseedot}. Next, Athos' float-to-fixed converter translates
the floating-point HLIL code to fixed-point code in a low-level
intermediate language LLIL. This step requires Athos to
compute the right precision to be used for maximum accuracy
(Section~\ref{subsec:athosquantizer}).
Figure \ref{fig:lrezpc} shows the
LLIL code snippet for logistic regression. The function calls in this
sequence can be implemented with a variety of secure computation
backends - e.g. ABY~\cite{aby} for the case of 2-party secure
computation, Porthos for the case of semi-honest 3-party secure
computation (Section \ref{sec:porthos}) and Aramis (Section
\ref{sec:aramis}) for the malicious secure variant. Different backends
provide different security guarantees and hence vary in their
performance. For this example, the three backends take
227ms, 6.5ms, and 10.2ms respectively.

\begin{figure}
\small
% \begin{minted}[mathescape,
%                linenos,
%                numbersep=5pt,
%                gobble=2,
%                frame=lines,
%                framesep=2mm]{python}
\begin{Verbatim}[frame=single]
# x is an MNIST image of shape (1,784).
# W and b are the model parameters.

print(tf.argmax(tf.matmul(x, W) + b, 1))
\end{Verbatim}
% \end{minted}
\caption{Logistic Regression: TensorFlow snippet}
\label{fig:lrtf}
\end{figure}

\begin{SaveVerbatim}{HLIL_LR_Verbatim}
xW = MatMul(x, W);
xWb = MatAdd(xW, b);
output(ArgMax(xWb));
\end{SaveVerbatim}

\begin{SaveVerbatim}[]{LLIL_LR_Verbatim}
//Assume Athos chooses
//15 bit precision

xW = MatMul(x, W);
ScaleDown(xW, 15);
xWb = MatAdd(xW, b);
output(ArgMax(xWb));
\end{SaveVerbatim}

\newcommand{\kw}[1]{{\ensuremath{\mathtt{#1}}}}
\newcommand{\ftext}[1]{\text{\small{#1}}}
\newcommand{\cond}[3]{\ensuremath{{{#1}\:?\:{#2}\::{#3}}}}
\newcommand{\forl}[4]{\ensuremath{\kw{for}\:{#1}\:\kw{in}\:[{#2}, {#3}]\:\kw{do}\:{#4}}}
\newcommand{\ite}[3]{\ensuremath{\kw{if}({#1}, {#2}, {#3})}}
\newcommand{\loops}[3]{\ensuremath{\kw{while}\:{#1} \leq {#2}\:\kw{do}\:{#3}}}

\section{Athos}
\label{sec:athos}
Athos compiles ML inference code written in \tensorflow to \mpc protocols. It has the following main components:

\begin{tiret}

\item \emph{Frontend.} Athos frontend compiles \tensorflow code to a
high-level intermediate language (HLIL). HLIL supports floating-point
tensors and sequence of function calls (corresponding to the
\tensorflow nodes) that manipulate tensors. The main challenge in the
frontend is to reconcile dynamic typing in \tensorflow to static
typing in HLIL. \tensorflow code, written in Python, does not have
tensor dimensions, whereas our HLIL has explicit tensor dimensions as
it enables the compiler to perform analyses and optimizations.

\item \emph{Float-to-fixed converter.} While ML models use
floating-point arithmetic, \mpc protocols operate on fixed-point
arithmetic. Rather than requiring the programmers to manually convert
(or re-train) their models to integers, Athos performs the conversion
automatically, without compromising on the inference accuracy.

\item \emph{Modular LLIL.} Athos compiles floating-point HLIL code to 
fixed-point code in a low-level
intermediate language (LLIL). LLIL is a C-like imperative language
that supports integer tensors, loops, conditionals, and functions. LLIL
also makes it easier for different cryptographic backends to be
plugged into Athos. It precisely specifies the interface that it
requires the cryptographic protocols to implement, while providing a
 library for other operations. The LLIL is compiled down to
the MPC protocol code.

\item \emph{Optimizations.} Athos implements 
 \mpc specific optimizations as well as
several standard dataflow
  analyses and compiler optimizations.
 The design of HLIL and LLIL, and the choice of them being
  statically typed, is partly motivated by the requirements of these
  analyses.

Below we explain each of these components in detail.
\end{tiret}
\subsection{Frontend and HLIL}
\label{subsec:athosfrontend}

Athos frontend compiles the input \tensorflow models to HLIL (described
next) with explicit tensor dimensions. To obtain these dimensions,
the frontend first runs \tensorflow code on one dummy input and
generates \tensorflow metadata that has all the required information.
The metadata is then translated to HLIL.

We discuss some details of the frontend.
A plain dump of  the \tensorflow metadata contains some nodes that 
are semantically irrelevant for actual inference,
e.g. {\tt identity}, {\tt assign}, etc. To avoid representing these
nodes in HLIL,
we first prune the \tensorflow graph to remove such nodes,
specifically we use the \tensorflow graph transform tool
\cite{tfGraphTransformTool} for this purpose. Next, Athos desugars the remaining (tens of) \tensorflow nodes  to HLIL, while keeping the number of functions in HLIL as small
as possible. \tensorflow also supports
``broadcasting'' \cite{tensorflowbroadcasting} that allows operations
on tensors of incompatible dimensions and sizes. For example, due to
broadcasting, addition of a four-dimensional tensor with a one-dimensional tensor is a valid operation. Athos frontend
passes the broadcasting information to HLIL,
which then accounts for it by compiling it to the appropriate LLIL
library function call.

% Finally, some \tensorflow graphs have ``control edges'' 
% \cite{abadi2017computational} that constrain the order of execution of
% the nodes -- Athos frontend also accounts for their mutability
% \aseem{Add detail how}.

\begin{figure}[htp]
  \footnotesize
  \[
  \begin{array}{rrcl}
    %\ftext{Function} & f &&\\
    \ftext{Constant} & n & ::= & 0 \mid 1 \mid 2 \mid \ldots\\
    \ftext{Float constant} & r & ::= & n.n\\
    \ftext{Type} & \hat{\tau} &::=& \kw{float} \mid \kw{int} \mid \hat{\tau}[n] \\
    \ftext{Matrix} & \hat{M} & ::= & \overline{r} \mid \overline{\hat{M}}\\
    \ftext{Expression} & \hat{e} &::=& n \mid x \mid \hat{M} \mid \hat{e_1} \oplus \hat{e_2} \mid x[\hat{e}] \\
    \ftext{Program} & \hat{p} & ::= &\kw{void}\ \kw{main}\;()\;\{\overline{\hat{\tau}\;x}\;;\overline{f(\overline{\hat{e}})}\}
  \end{array}
  \]
\caption{HLIL syntax}
\label{fig:hil}
\end{figure}

Figure~\ref{fig:hil} shows the HLIL (we use $\overline{r}$ to denote
sequences of floating-point constants, and similarly for other
syntactic categories). It is a simple language of floating-point
tensors ($\hat{M}$), with dimensions ($n$) and sizes as explicit type annotations ($\hat{\tau}[n]$), and
the $\kw{main}$ is a sequence of variable declarations and function calls.

We next discuss how Athos performs float-to-fixed conversion on HLIL
programs.

\subsection{Float-to-fixed}
\label{subsec:athosquantizer}
As observed earlier, most ML models are expressed using
floating-point, while \mpc protocols operate on
integers. For large models, we cannot expect the programmers to
manually translate or re-train floating-point ML models to integer
code (the common approach in literature on secure inference~\cite{secureml,minionn,gazelle,aby3,securenn,delphi,chameleon}).
Furthermore, it is well-known that floating-point operations are much
more inefficient than fixed-point when evaluated
securely~(\cite{secureml,aby3}) -- we re-confirm this by performing
two-party secure multiplication~\cite{ddkssz15} using both fixed-point
and floating-point arithmetic to showcase the difference. This is
illustrated in Table \ref{tab:floatvsfixed} which shows the huge
overheads associated with floating-point arithmetic.
In future, if efficient protocols for floating-point become
available then we can directly compile HLIL to them, but until then
Athos automatically performs the translation.

The translation is parametrized by a scale parameter
$s$ that determines the precision.
We discuss how this scale is set later in the section. Given a scale
$s\in\mathbb{Z}$, we define a map $\rho_s:\mathbb{R}\rightarrow \mathbb{Z}_{2^b}$
that maps Reals to $b$-bit integers: $\rho_s(r)=\lfloor r\cdot2^s\rfloor$. We abuse notation and also apply $\rho_s$ to
matrices $M$ of Reals where the result is a point-wise application of
$\rho_s$ to each matrix element.  
In the output fixed-point code, every Real number $r$ is represented by a $b$-bit integer.
The Real representation of an integer $n$ is given by $\frac{n}{2^s}$.
%the conversion algorithm
The float-to-fixed conversion (for select cases) is described in
the following algorithm (\kw{ScaleDown} is described in
Table~\ref{tab:smf}):

\[
\begin{array}{lcl}
%F(n) & = & n\\
%F(r) & = & \rho_s(r)\\
%F(\hat{M}) & = & \rho_s(\hat{M})\\
%F(\kw{float}\ x) & = & \kw{int}\ x\\
F(\kw{MatAdd}(A,B,C)) & = & \kw{MatAdd}(A,B,C)\\
F(\kw{MatMul}(A,B,C)) &= &\kw{MatMul}(A,B,C);\\
& &\kw{ScaleDown}(C,s)
\end{array}
\]

%Here, $\mathit{unop}$ denotes a generic unary operator (e.g., $\mathtt{argmax},\mathtt{maxpool},\mathtt{relu},\dots$).
%In the first step, we run the a procedure that replaces all floating-point numbers in the program with fixed-point integers. 
% The only operation which needs to change because of float-to-fixed is matrix multiplication. Since convolution calls matrix multiplication (Figure~\ref{convtomatmul}), it needs to be modified to call the new matrix multiplication (Figure~\ref{newmatmul}).

%example
\
As an example of the conversion process, consider the program $M_1*M_2$ that multiplies
the row vector $M_1=[400.1,200.1]$ with the column vector $M_2=[0.3,0.1]^T$.
Then in infinite precision Real arithmetic the result of the computation
$400.1*0.3+200.1*0.1$
is $140.04$. Single-precision floating-point 
arithmetic with 32 bits only has a 23-bit mantissa and computes the approximately correct result 140.040009.
%When using Athos, the computed result
%can be much more precise than the floating-point result. 
We use $0.1f$ to denote the floating-point number closest to the Real number $0.1$. 
Given $s=24$, $F(M_1*M_2)$ results 
into the following program over integers
\[
(\rho_{24}(400.1f)*\rho_{24}(0.3f)+\rho_{24}(200.1f)*\rho_{24}(0.1f)) >> {24}
\]
which results in the following computation with 64-bit integers
\[
(6712564224*3357121024+5033165*1677721)>>{24}
\]
The final result is 2349481329 that represents the real number
$\frac{2349481329}{2^{24}}=140.040000021457672119140625$ which is good approximation of the desired result $140.04$. Although it is feasible to constuct examples where fixed-point computations
can be imprecise, ML usually operates on normalized values and we have observed that Athos does not lose accuracy in practice (Table~\ref{tab:fixed-accuracy}).

Athos, assigns the same bit-width $b$ and the same scale $s$ to all
network parameters. While we could use different $b$ and $s$, our
experimental results show that same values for all parameters works
quite well in practice.
%Hence, no information about any individual parameter gets leaked. 
We keep the scale public for efficiency: division with $2^s$ when $s$ is secret is much more expensive than when $s$ is public.
Moreover, scaling down operations (division by $2^s$) cause loss of precision, as they lose significant bits, and hence need to be minimized.
 Therefore, Athos scales down only once per matrix multiplication and does not scale down matrix additions.
%These design choices make Athos ``crypto-aware" and effective for secure machine learning (Section~\ref{subsec:athosexperiments}).
%Prior float-to-fixed converters assign different bit-widths/scales to different parameters and  scale down after every operation~\cite{}.
%Because IEEE754 floating-point numbers have a 23-bit mantissa, if $s$ is set to 23 and all intermediate values occurring at runtime are below 128 then the fixed-point code has very similar accuracy to floating-point code. 
%However, in practice, the intermediate values do exceed 128, causing $s$ being set to smaller values, and some accuracy is lost.

%\subsubsection{Setting Scale}\label{subsec:athossettingscale}
%We try all possible $s$. If $s$ is large then overflows and garbage. If $s$ is small then imprecise %result. Autotuning gives a good $s$ that produces good results. 

While we use machine integer width (64) for $b$, finding a good value
of $s$ is difficult. We explain the various tradeoffs that
govern the choice of $s$ and then discuss our solution.

Suppose, in our example, $s$ is set too low: $s=2$.
Then $F([400.1f,200.1f]*[0.3f,0.1f])$ is $(1600*1+800*0)>>2$,
which represents the Real number $400/4=100$.
This result is far from 140.04. Here, low scale values have lead to
loss of significant bits. In particular, 0.1 has been rounded to zero
causing an imprecise result. Ideally we want to set the scale to a large value
so that the integers have many significant digits.

Next, suppose $s$ is set to a very high value, e.g., 60. Then, the
computation $\rho_{60}(400.1f)*\rho_{60}(0.3f)$ overflows 64-bit integers and
the result is garbage
(multiplication of these two large positive numbers would become a negative number).  
%Usually, at compile time the operations are of the form $x*y$ where the values of $x$
%and $y$ are known only at runtime. Hence, it is hard to decide whether a particular
%scale value would lead to overflows or not at compile time.

Thus, scale can neither be very low nor very high; we need to find a sweet spot.
To determine an appropriate value of $s$, we sweep over all its possible values $\{0,1,\ldots,b-1\}$
and choose the value that leads to the best accuracy. For the example $400.1f*0.3f+200.1f*0.1f$,
the most accurate result is obtained at $s=24$. In general, machine learning algorithms have a 
 validation dataset that is used for hyperparameter tuning. We consider scale as a hyperparameter
and select the scale that leads to a fixed-point classifier implementation that performs the best on the validation set.
The scale chosen by Athos is a {\em leakage} function that depends on the weights of the model.
Athos gives a methodical way of picking this scale that prior works did manually.
Hence, leakage by Athos is similar to all prior works on secure inference. 
%We point out that while the scale chosen by Athos is a ``leakage'' function that depends on the weights of the model, we stress that this is not unique to our work. Indeed, all prior works on secure evaluation of machine learning algorithms pick an appropriate scale with the hope that the accuracy of the fixed-point network will not be too much worse than the original floating-point version. Our scheme helps us select the scale that helps match the floating-point accuracy.

%Like all prior works on secure inference, we keep the scale public for efficiency reasons.
%Hence, the {\em{leakage}} by Athos is identical to prior work.

%The accuracy of the final classifier is reported on the test set. In particular, the test set
%is used only for evaluation and {\em not} to generate a classifier implementation. Athos only works with the validation
%set and does not have access to the test set. 

\begin{table}
  \centering
  \resizebox{\columnwidth}{!}{

      \begin{tabular}{|c|c|c|c|l|}
    \hline
    \# Sequential Multiplications & Fixed (ms) & Float (ms) & Overhead \\
    \hline
  $1$ & $2.57$ & $72.35$ & $28.11$x\\
  \hline
    $10$ & $4.88$ & $278.8$ & $57.1$x \\  \hline
    $100$ & $21.65$ & $2735$ & $126.34$x \\   \hline
    $1000$ & $199.6$ & $25281.42$ & $126.6$x \\   \hline
\end{tabular}
}
 \caption{Floating-point vs Fixed-point multiplication.}
\label{tab:floatvsfixed}
%\tableup
\end{table}

\subsection{Modular LLIL}
\label{sec:athosmodularity}

\begin{figure}[htp]
  \footnotesize
  \[
  \begin{array}{rrcl}
    \ftext{Constant} & n & ::= & 0 \mid 1 \mid 2 \mid \ldots\\
    \ftext{Type} & \tau &::=& \kw{int} \mid \tau[n] \\
    \ftext{Matrix} & M & ::= & \overline{n} \mid \overline{M}\\
    \ftext{Expression} & e &::=& n \mid x \mid M \mid e_{1} \oplus e_{2} \mid x[e]\\
    \ftext{Statement} & s &::=& \tau\:x \mid x = e \mid \kw{for}(x=e_{1};x<e_{2};x++)\{s\}\\
    & & \mid& x[e_{1}] = e_{2} \mid \kw{if}(e, s_{1}, s_2\} \mid
    s_{1}; s_{2} \mid \kw{return}\ e\\
    & & \mid & f\;(\overline{e}) \mid d\;(\overline{e})\\
    \ftext{Global} & g & ::= &\kw{extern}\;\tau\;d\;(\overline{\tau\;x}) \mid \tau\;f\;(\overline{\tau\;x})\{s\}\\
    \ftext{Program} & p & ::= & \overline{g};\;\kw{void}\ \kw{main}\;()\;\{s\}
  \end{array}
  \]
\caption{LLIL syntax}
\label{fig:lil}
\end{figure}

Athos compiles HLIL to LLIL, a crypto-aware, C-like intermediate
language that has only integer-valued tensors. Figure~\ref{fig:lil}
shows the syntax of LLIL. This language  has sufficient expressiveness
required to implement ML inference tasks. In particular it supports
arrays, basic arithmetic, loops, branching, functions, and \kw{extern}
declarations.
LLIL makes the Athos interface to the \mpc cryptographic protocols
explicit. We observe that the tensor operations in a typical
\tensorflow code fall into two categories: those that do not change
the values but just copy the data around (e.g. {\tt squeeze} to remove
dimensions of size 1 from a tensor, {\tt pad} to pad a tensor with
various kinds of paddings, {\tt transpose} to take the transpose of a
tensor, and {\tt concat} to concatenate two tensors into a single
tensor), and those that compute new values.
For functions that do not manipulate shares (denoted by $f$), LLIL
provides a library with their
implementations that is automatically added as a prelude to LLIL
programs. Changing the underlying crypto protocol does not require
changes to these library functions and this library can be used by all
crypto developers. These functions are implemented in LLIL and are
compiled to C++ code.

Share-manipulating functions ($\kw{extern}\;d$) are required to be
implemented in the cryptographic backend.
All a crypto developer needs to do is to implement these functions,
and then she would be able to directly evaluate
the protocols on ML models used in practice. We describe these
functions with their signatures and intended semantics in Table~\ref{tab:smf}.
Concretely, we provide three implementations of these functions: using the 2PC
protocols of ABY~\cite{aby}, 3PC protocols of SecureNN~\cite{securenn},
and Porthos (Section~\ref{sec:porthos}). 

\begin{table*}
\begin{tabular}{rl}
MatMul(int[L][M] A, int[M][N] B, int[L][N] C) & Multiply two tensors $A$ and $B$ and store results in $C$\\
MatAdd(int[L][M] A, int[L][M] B, int[L][M] C)& Add two tensors $A$ and $B$ into $C$\\
Conv(int[H][W][CI] A,  int[FH][FW][CI][CO] F) & Convolve a tensor $A$ with filter $F$\\
% Optional parameters include padding and strides. \\
Avg/Max Pool(a, b, int[H][W][C] A) & Apply a stencil that computes the average/max value in windows of size $a\times b$ of tensor $A$.\\
ArgMax(int[M] A) & Compute the index with maximum value in $A$ \\
FusedBatchNorm(int[K][L][M][N] A, int[N] B, int[N] C) & Returns $\forall k,l,m,n. B[n] \times A[k][l][m][n] + C[n]$\\
ReLU(int[M][N] A) & Returns $\forall i,j. {\sf Max}(A[i][j],0)$\\
ScaleDown(int[M][N] A, k) & Arithmetic right shift each entry of $A$ with $k$.\\
\end{tabular}
\caption{Share manipulating functions. These have been simplified for
  exposition by suppressing parameters such as padding and
  strides. For comprehensive signatures, see~\url{https://www.tensorflow.org/api_docs/python/tf/}.}
\label{tab:smf}
\end{table*}

Finally, Athos compiles LLIL programs to C++ and links them with
the cryptographic \mpc protocol implementation.

\subsection{Optimizations}
\label{sec:athosopt}
Athos intermediate languages are designed to be amenable to static analysis. 
In particular, we have implemented several standard dataflow analyses and compiler optimizations~\cite{dragonbook}:
reaching definitions, liveness analysis, and so on. These analyses help with
optimizing memory utilization and we have observed
savings reaching up to 80\%. 
To demonstrate the ease of implementing analyses and optimizations, we provide an example each:
(a) a peephole optimization ReLU MaxPool Switching on HLIL to improve
efficiency of DNNs that use ReLU and MaxPool, and (b) an analysis
Counting Scale Down operations on LLIL to determine the number of scale
down operations done in order to prevent loss in
accuracy (a similar analysis was done manually in~\cite{secureml,securenn,aby3}).

\subsubsection{ReLU MaxPool Switching}
% The ReLU operation is one of the most time intensive task in secure inference of DNNs. For some DNNs, secure evaluation of ReLUs can consume up to 80\% of the total protocol execution %time. This is in contrast to evaluation in the clear where ReLUs consume only a fraction of the total time.
%Hence, it is plausible that ML developers can write \tensorflow code
%in a way that has no impact on cleartext evaluation  but can severely
%degrade the performance of secure evaluation. One such idiom involves
%applying ReLU to a matrix followed by MaxPool. Notice that ReLU and
%MaxPool are commutative operators: ReLU(MaxPool($\cdot$)) is
%functionally equivalent to MaxPool(ReLU($\cdot$)). Moreover, for
%cleartext performance, there is no discernible difference in the
%performance of these two alternatives. Hence,
Most \tensorflow
developers have adopted the convention of expressing DNN layers using the MaxPool(ReLU($\cdot$)) idiom.
For protocols like Porthos and SecureNN~\cite{securenn} that reduce ReLU and MaxPool to secure comparison protocols, ReLU(MaxPool($\cdot$)) can be much more efficient than  MaxPool(ReLU($\cdot$))  as this significantly reduces the number of comparisons. As opposed to SecureNN, where this was done manually, we have built a peephole optimization pass on HLIL
that replaces occurrences of $\kw{MaxPool}(a,b, \kw{ReLU}(A));$ with $\kw{ReLU}(\kw{MaxPool}(a,b,A));$. For example,
if the input matrix $A$ has dimensions $112\times112\times64$ and we compute a MaxPool with $2\times2$ windows.
Then, the output matrix has dimensions $56\times56\times 64$. Hence, the latter needs to compute only one fourth the number of ReLUs
compared to the former. In this case, the optimized code is over $3\times$ better in communication and over $2\times$ faster in our experimental setup (Section~\ref{sec:experiments}).
\subsubsection{Counting Scale Down operations}
We describe an analysis to count the number of scale down operations in an LLIL code. The analysis uses an environment $\rho$ that
maps tensors to the number of elements they contain. This
environment is populated using variable declarations in the code.
 The analysis makes a single pass over $\kw{main}$ and for each call
 $\kw{ScaleDown(A,s)}$ 
accumulates $\rho(A)$ into a counter. The final value of the counter
provides the number of scale down operations in the code.

Note that this analysis is easy to describe as the LLIL code contains
dimensions of all the tensors explicitly.
Hence, the compiler can statically populate $\rho$. This analysis is
impossible to perform on the \tensorflow Python code 
as the sizes of tensors are unknown at compile time.

\section{Porthos}\label{sec:porthos}

We now describe Porthos, our improved secure 3PC protocol that
provides semi-honest security against one corrupted party
and privacy against one malicious corruption. The notion of privacy
against malicious corruption (introduced by Araki
\etal~\cite{maliciousprivacy}) informally guarantees that privacy of
inputs hold even against malicious party as long as none of the parties participating in the
protocol learn the output of the computation (this is relevant, for
example, when computation is offloaded to servers). Porthos builds
upon SecureNN~\cite{securenn} but makes crucial modifications to
reduce communication.
% and computation overheads. %, however we only rely on its full security against one semi-honest corruption here. 
We first describe our protocols that reduce communication and
summarize concrete improvements in Table~\ref{tab:protcomplexity}.
%We describe the compute optimizations of Porthos in Section \ref{subsec:porthoscomp}.
% In Section \ref{subsec:porthosexperiments}, we show our experimental results illustrating how Porthos outperforms relevant prior works.
% such as SecureNN~\cite{securenn}, ABY3~\cite{aby3}, EzPC~\cite{ezpc}, and CHET~\cite{chet}.

%\subsection{Reducing Communication}\label{subsec:porthoscomm}
%\dg{Need to polish this para.}
We reduce communication for both linear as well as non-linear layers
of DNNs. Linear layers  include fully connected layers as well as
convolutional layers.
We improve the communication for convolutional layers and our optimization gains get better with larger filter sizes.
%We modify how convolutional layers are computed in SecureNN. Our improvements are more pronounced when the network uses large filters in the convolution. 
With regards to non-linear layers (ReLU and MaxPool), we modify how
two of the protocols in SecureNN are used -- ComputeMSB and
ShareConvert. 
As we explain below, this directly translates to better communication for both ReLU and MaxPool computations. 
At a very high level, we trade communication with compute by modifying
the way certain shares are generated in the protocol.
\\\\
\noindent {\bf Convolution.} In SecureNN, secure computation of
convolutional layers is done by reducing them to a (larger) matrix
multiplication. As an example,  $2$-dimensional convolution of a
$3\times 3$ input matrix $X$ (with single input channel and stride 1)
with a filter $Y$ of size $2 \times 2$ reduces to a matrix
multiplication as follows:
%\vspace{-0.19in}
% \[ 
%\begin{small}
\begin{flalign*}
% \scriptsize{
&\mathsf{Conv2d}\left(\begin{bmatrix}
    x_{1}       & x_{2} & x_{3} \\
    x_{4}       & x_{5} & x_{6} \\
    x_{7}       & x_{8} & x_{9}
\end{bmatrix},
\begin{bmatrix}
    y_{1}       & y_{2} \\
    y_{3}       & y_{4}
\end{bmatrix}\right )
% }
= & 
% \]
\end{flalign*}
\[ %\hspace{-0.1cm} 
% \scriptsize{
\hspace{100pt}
\begin{bmatrix}
    x_{1}       & x_{2} & x_{4} & x_5 \\
    x_{2}       & x_{3} & x_{5} & x_{6}\\
    x_{4}       & x_{5} & x_{7} & x_{8}\\
    x_{5}       & x_{6} & x_{8} & x_{9}\\
\end{bmatrix}\times
\begin{bmatrix}
    y_{1}       \\
    y_2 \\
    y_3 \\
	y_4 \\
\end{bmatrix}
% \hfilneg
% }
\]
%\end{small}
%\vspace{-0.15in}
In the above matrix multiplication, we call the left matrix (derived from $X$) as the ``reshaped input'' (say, $X'$) and the right matrix (derived from $Y$) as the ``reshaped filter'' (say, $Y'$). 
%This can be generalized in a similar manner to other dimensions and parameters (such as padding, stride etc.) - see e.g.~\cite{convnotes}. 
The matrix multiplication  is computed securely using a matrix Beaver triple~\cite{beaver,secureml} based protocol. Later, the output can be reshaped to get the output of convolution in correct shape.
In this protocol, matrices being multiplied are masked by random matrices of same size and communicated and hence, the communication grows with the size of the matrices. 
We observe that this is quite wasteful for convolution because the reshaped input image (the first matrix in multiplication) has many duplicated entries (e.g., $x_2$ in row 1 and row 2) that get masked by independent random values. 
%For instance, value $x_2$ in row 1 and row 2 would be masked by independent random values and communicated
Let size of $X$ be $m\times m$ and size of $Y$ be $f\times f$. Then, the size of $X'$ is $q^2\times f^2$, where $q = m-f+1$.
%However, doing this is ``wasteful'' in the number of Beaver triples. To see this, in the above example, when the convolution is expressed as a matrix multiplication on the right side, the variable $x_2$ in row $2$ is treated as a ``new'' variable, different from the variable $x_2$ in row $1$. Hence, a ``fresh'' Beaver triple is utilized when computing the product of this $x_2$ with $k_1$. This can be avoided with the knowledge of the structure of the matrix multiplication. 
In Porthos, we optimize the size of matrix-based Beaver triples for convolution by exploiting the structure of re-use of elements as the filter moves across the image. 
At a high level, we pick random matrix of size matching $X$ for masking and communication only grows with size of $X$ (i.e., $m^2$) instead of $X'$ (i.e., $q^2f^2$) in SecureNN.

Before, we describe our optimized protocol, we set up some notation. Let $\share{x}{t}{0}$ and $\share{x}{t}{1}$ denote the two shares of a 2-out-of-2 additive secret sharing of $x$ over $\bbZ_t$ -- in more detail, pick $r \atrand \bbZ_t$, set $\share{x}{t}{0} = r$ and $\share{x}{t}{1} = x-r\ (\mathsf{mod}\ t)$. $\shareval{x}{t}$ denotes a sharing of $x$ over $\bbZ_t$. 
%The algorithm $\genshare{t}{x}$ generates the two shares of $x$ over $\bbZ_t$ and algorithm $\reconst{t}{x_0, x_1}$ reconstructs a value $x$ using $x_0$ and $x_1$ as the two shares (reconstruction is simply $x_0+x_1$ over $\bbZ_t$). 
Reconstruction of a value $x$ from its shares $x_0$ and $x_1$ is simply  $x_0+x_1$ over $\bbZ_t$.
This generalizes to larger dimensions - e.g. for the $m\times n$ matrix $X$, $\share{X}{t}{0}$ and $\share{X}{t}{1}$ denote the matrices that are created by secret sharing the elements of $X$ component-wise (other matrix notation such as $\reconst{t}{X_0,X_1}$ are similarly defined). 

Let $\mathsf{Conv2d}_{m,f}$ denote a convolutional layer with input $m\times m$, $1$ input channel, a filter of size $f\times f$, and $1$ output channel. 
Our protocol for $\mathsf{Conv2d}_{m,f}$ is described in Algorithm~\ref{algo:conv2d}, where $L = 2^\ell$, $\ell=64$. Algorithms  $\mathsf{ReshapeInput}$, $\mathsf{ReshapeFilter}$ and $\mathsf{ReshapeOutput}$ are used to reshape input, filter and output as described above and are formally described in Appendix \ref{appendix:porthos}. 
Parties $P_0$ and $P_1$ start with shares of input matrix $X$ and filter $Y$ over $\bbZ_L$ That is, $P_j$ holds $(\share{X}{L}{j}, \share{Y}{L}{j})$ for $j \in \{0,1\}$.
In SecureNN, $P_0$ first reshapes $\share{X}{L}{0}$ into $\share{X'}{L}{0}$ by running $\mathsf{ReshapeInput}$. Then, it picks a random matrix $\share{A'}{L}{0}$ of same size as $X'$ and sends $\share{E'}{L}{0} = \share{X'}{L}{0} - \share{A'}{L}{0}$ to $P_1$ that requires communicating $q^2f^2$ elements.
In Porthos, we optimize this as follows: $P_0$ picks a random matrix $\share{A}{L}{0}$ of same size as $X$ (Step 1) and sends $\share{E}{L}{0} = \share{X}{L}{0} - \share{A}{L}{0}$ to $P_1$ (Step 4) that requires communicating $m^2$ elements only. Later, parties can reshape $E$ locally to get $E'$. 
We reduce the communication by $P_1$ in a symmetric manner.
Concretely, we reduce communication from $(2q^2f^2+2f^2+q^2)\ell$ in SecureNN to $(2m^2+2f^2+q^2)\ell$.
This algorithm can  be easily generalized to the setting where there are $i$ input filters, $o$ output filters, and different stride and padding parameters. 

\begin{algorithm}
\KwIn{$P_0$  holds $(\share{X}{L}{0}, \share{Y}{L}{0})$ and $P_1$ holds $(\share{X}{L}{1}, \share{Y}{L}{1})$, where $X \in \bbZ_L^{m \times m}$, $Y \in \bbZ_L^{f \times f}$.}
\KwOut{$P_0$ gets $\share{\mathsf{Conv2d}_{m,f}(X,Y)}{L}{0}$ and $P_1$ gets $\share{\mathsf{Conv2d}_{m,f}(X,Y)}{L}{1}$.}
\textbf{Common Randomness}: $P_0$ \& $P_1$ hold shares 
%$\share{U}{L}{j}$, $j \in \{0,1\}$, 
of a zero matrix $U$ of dimension $q \times q$, $q = m-f+1$ . $P_0$ \& $P_2$ hold a common PRF key $k_0$, and $P_1$ \& $P_2$ hold a common PRF key $k_1$.
\begin{enumerate}
\item $P_0$ \& $P_2$ use PRF key $k_0$ to generate random matrices 
%$\share{A}{L}{0}$, $\share{B}{L}{0}$ and $\share{C}{L}{0}$ where 
$\share{A}{L}{0} \in \bbZ_L^{m \times m}$, $\share{B}{L}{0} \in \bbZ_L^{f \times f}$ and $\share{C}{L}{0} \in \bbZ_L^{q\times q}$.
\item $P_1$ \& $P_2$ use PRF key $k_1$ to generate random matrices 
%$\share{A}{L}{1}$ and $\share{B}{L}{1}$ where 
$\share{A}{L}{1} \in \mathbb{Z}_L^{m \times m}$ and  $\share{B}{L}{1} \in \mathbb{Z}_L^{f \times f}$.
%\end{itemize}
\item $P_2$ computes $A = \share{A}{L}{0} + \share{A}{L}{1}$ and $B = \share{B}{L}{0} + \share{B}{L}{1}$. 
Let
%\item $P_2$ now computes 
$A' = \mathsf{ReshapeInput}(A)$ and $B' = \mathsf{ReshapeFilter}(B)$. $P_2$ computes $\share{C}{L}{1} = A' \cdot B' - \share{C}{L}{0}$ and sends it to $P_1$.
\item For $j \in \{0,1\}$, $P_j$ computes $\share{E}{L}{j} = \share{X}{L}{j} - \share{A}{L}{j}$ and $\share{F}{L}{j} = \share{Y}{L}{j} - \share{B}{L}{j}$ and sends to $P_{j\oplus 1}$.
\item $P_0$ \& $P_1$ reconstruct $E$ and $F$ using exchanged shares.
\item For $j \in \{0,1\}$, $P_j$ computes $\share{X'}{L}{j} = \mathsf{ReshapeInput}(\share{X}{L}{j})$, $E' = \mathsf{ReshapeInput}(E)$,  $\share{Y'}{L}{j} = \mathsf{ReshapeFilter}(\share{Y}{L}{j})$, $F' = \mathsf{ReshapeFilter}(F)$.
%\item For $j \in \{0,1\}$, $P_j$ computes $\share{X'}{L}{j} = \mathsf{ReshapeInput}(\share{X}{L}{j})$ and $\share{Y'}{L}{j} = \mathsf{ReshapeFilter}(\share{Y}{L}{j})$, followed by $E' = \mathsf{ReshapeInput}(E)$ and $F' = \mathsf{ReshapeFilter}(F)$.
\item For $j \in \{0,1\}$, $P_j$ computes $\share{Z'}{L}{j} = -jE' \cdot F' + \share{X'}{L}{j} \cdot F' + E' \cdot \share{Y'}{L}{j} + \share{C}{L}{j} + \share{U}{L}{j}$.
\item For $j \in \{0,1\}$, $P_j$ outputs $\share{Z}{L}{j} = \mathsf{ReshapeOutput}(\share{Z'}{L}{j})$.
\end{enumerate}
    \caption{{3PC protocol for $\mathsf{Conv2d}_{m,f}$ } \label{algo:conv2d}}

\end{algorithm}

\noindent {\bf Activation Functions.} 
In SecureNN protocols for computing activations such as ReLU and MaxPool start with parties $P_0$ and $P_1$ having shares of values over $L = 2^{64}$. 
For both of these, parties run a protocol called $\computemsb$ to evaluate most significant bit (MSB) of secret values. This protocol require shares over $L-1$. So parties run a protocol called $\shareconvert$ to convert shares over $L$ to shares over $L-1$. Both protocols $\computemsb$ and $\shareconvert$ require $P_2$ to send fresh shares of a value to $P_0$ and $P_1$. In SecureNN, both of these shares were picked by $P_2$ and explicitly communicated to $P_0$ and $P_1$.  As mentioned before, shares of a value $x$ are $r$ and $x - r$, where $r$ is a appropriately picked uniformly random value. We observe that since one of the shares is truly random, it can be computed as the output of a shared PRF key between $P_2$ and one of the parties, say $P_0$. This cuts the communication of this step to {\em half}. Moreover, since many activations are computed in parallel, we can carefully ``load-balance'' this optimization between $P_0$ and $P_1$ to reduce the communication to half on the critical path. We implement this load-balance optimization and observe that this reduces the overall communication of $\shareconvert$, $\computemsb$, $\relu$ and $\maxpool$ by $25\%$. 

%When computing activation functions such as $\relu(x)$ and their derivatives, SecureNN goes through a series of protocols where parties $P_0$ and $P_1$ start with additive 2-out-of-2 shares of $x$ over $\bbZ_L$. They move to shares of the same $x$ over $\bbZ_{L-1}$ (with the help of $P_2$) using a protocol called $\shareconvert$, and then convert the problem of computing $\relu$ to computing the LSB of $x$ when $x$ is shared over $\bbZ_{L-1}$, using another protocol called $\computemsb$. In both $\shareconvert$ and $\computemsb$ protocols, $P_2$ must generate fresh shares of a value that is reconstructed in the protocol and send them to $P_0$ and $P_1$. In SecureNN, both these shares were computed by $P_2$ and sent to $P_0$ and $P_1$. We observe that since one of these shares is truly random, they can be computed through a PRF key shared between $P_2$ and (say) $P_0$, and hence only one of these shares must be communicated to (say) $P_1$. This roughly halves the communication of these protocols. Additionally, since these protocols are called in parallel several times, in order to ``load balance'' the communication, we have $P_2$ set the share to be given to $P_0$ as a PRF output in half of the invocations and the share to be given to $P_1$ as a PRF output in the other half of the invocations. 

The revised table with comparison of overall communication complexity of all protocols with improvements over SecureNN are provided in Table \ref{tab:protcomplexity}. 
%Only protocols whose communication complexity are an improvement over SecureNN are presented. 
$\mathsf{Conv2d}_{m,i,f,o}$ denotes a convolutional layer with input $m\times m$, $i$ input channels, a filter of size $f\times f$, and $o$ output channels. 
%$\drelu$ denotes the derivative of $\relu$. 
$\maxpool_n$ computes the maximum value out of a list of $n$ elements. $\prm$ denotes a prime value strictly larger than $65$ (set to $67$ in SecureNN), with $8$ bits being used to represent elements in $\bbZ_\prm$ (hence $\log \prm = 8$ in the table). %The round complexity of none of the protocols change and remain the same as in SecureNN.

\begin{table}
  \centering
  \resizebox{\columnwidth}{!}{
      \begin{tabular}{|l|l|l|l|}
    \hline
    Protocol & Communication (SecureNN) & Communication (Porthos) \\    \hline
 $\mathsf{Conv2d}_{m,i,f,o}$ & $(2q^2f^2i+2f^2oi+q^2o)\ell$ & $(2m^2i+2f^2oi+q^2o)\ell$\\ \hline
	$\shareconvert$ & $4\ell\log\prm+6\ell$ & $3\ell\log\prm+5\ell$ \\ \hline
    $\computemsb$ & $4\ell\log\prm+13\ell$ & $3\ell\log\prm+9\ell$ \\ \hline

$\relu$ & $8\ell\log\prm+24\ell$ & $6\ell\log\prm+19\ell$  \\ \hline
       $\maxpool_n$ & $(8\ell\log\prm+29\ell)(n-1)$ & $(6\ell\log\prm+24\ell)(n-1)$\\ \hline

\end{tabular}
}
 \caption{Communication complexity of protocols; $q = m-f+1$ and $\log \prm = 8$.}
\label{tab:protcomplexity}
%\tableup
	% \vspace{-0.8cm}
\end{table}

\section{Aramis}\label{sec:aramis}
%\vspace{-0.1cm}

In this section, we describe Aramis, a general technique to convert
any semi-honest secure MPC
protocol into a secure MPC protocol tolerating malicious
corruptions by relying on secure hardware. The threshold of corrupted parties tolerated by the
semi-honest protocol is retained in the malicious
secure protocol by our technique. 
\\\\
\noindent\textbf{Threat Model.} We consider a strong threat model where not only does the adversary control the operating system of the corrupted parties 
(i.e., the host operating system is outside the Trusted Computing Base)
but also observes the entire state of their secure hardware.
Aramis makes a very minimal
trust assumption of {\em integrity} on
hardware, namely that of code attestation (the outputs generated by the hardware are indeed from the code that it attested to).
This implicitly requires
the hardware to possess a trusted component that can produce
signatures and this signature scheme cannot be forged
by the adversary. 
However, the adversary can see the state  (i.e., all the code and the user data) of the hardware belonging to the corrupted parties, i.e., we {\em do not} assume {\em confidentiality} of state.
Prior works~\cite{vc3, obliviousmpml, GuptaFC16, BahmaniFC17,gcsgx,ndss1,ndss2,ndss3,slalom,opaque,chiron} that combine \mpc and hardware (SGX)  make stronger trust assumption on the hardware of both confidentiality and integrity,
 and hence, provide security only in a weaker threat model where the hardware hides the data residing in it from the adversary. 
%Aramis makes a very minimal
%trust assumption of {\em integrity} on
%hardware: the code and data residing in the hardware cannot be modified by the adversary.
%This implicitly requires
%the hardware to possess a trusted component that can produce
%signatures and this signature scheme cannot be forged
%by an adversary. 
%We do not assume confidentiality, that is, the adversary can see all the code and user data that resides in  the hardware belonging to the corrupted parties. 
%This significantly weakens the trust assumption on hardware compared to prior work that combines \mpc and SGX (\cite{vc3, obliviousmpml, GuptaFC16, BahmaniFC17,gcsgx,ndss1,ndss2,ndss3,slalom}).
\\\\
\noindent\textbf{Overview.} At a very high level, Aramis
exploits the following (well-known) observation: in order for a
semi-honest protocol to be made maliciously secure, one must ensure
that all messages sent by every party $P_i$ are computed honestly
according to the specification of the semi-honest protocol consistent
with $P_i$'s input and the transcript so far.
The next observation we make is that if party $P_i$ possesses hardware
whose code can be attested by party $P_j$ (and vice-versa), then $P_j$
can obtain guarantees on the correctness of protocol messages sent by
$P_i$ as long as these messages are computed and signed by $P_i$'s
hardware.
Using these observations, we can convert a semi-honest secure protocol into one that is maliciously secure by having every protocol message of $P_i$ be computed by the trusted hardware that $P_i$ executes. 
We shall now describe our techniques in more detail. 
We first describe the ideal functionality that is assumed out of the hardware in Section \ref{subsec:sgxideal}. 
We then describe our technique in Section~\ref{sec:shtomcompiler}.
Finally, we provide an implementation of Aramis using Intel SGX as the underlying secure hardware. 
We explain how Intel SGX can realize the ideal functionality
in Section~\ref{sec:fattest-sgx} and challenges in porting semi-honest
\mpc protocols to SGX in Section~\ref{sec:challenges-aramis}.
%The structure of the underlying semi-honest secure MPC protocol that we use is formalized in Section \ref{subsec:nextmessagefunction} and we present our compiler in Section \ref{subsec:shtomcompiler}. Finally, we describe some of the challenges in porting secure computation protocols to SGX in Section \ref{subsec:porting}.
%\vspace{-0.23cm}

\subsection{The attestation ideal functionality $\fattest$}
\label{subsec:sgxideal}
\noindent\textbf{Description.} We formally define the ideal functionality for attested executions 
%of deterministic functions 
in Figure \ref{fig:attestideal}.
% and later describe in Section~\ref{sgxtoideal} how Intel SGX can be used to realize this functionality.
The functionality is parameterized by a signing key pair $(\vk,\sksign)$. Let 
 $\sign_{\sksign}(m)$ denote the signing algorithm on message $m$ and $\verify_{\vk}(m,\sigma)$ denote verification of signature $\sigma$ on message $m$. 
At a high level, this functionality allows users to specify a function $g$ to the ideal functionality once using the $\gcommit$ command. 
The functionality returns a token $\token_g$ generated as $\sign_\sksign(H(g))$, where $H$ is a collision resistant hash function. 
Note that this token is publicly verifiable given $g$ and $\vk$.
Let $\st_\ctr$ be an internal state that the functionality maintains,
indexed by $\ctr$ -- this state can be maintained by signing it along
with $\ctr$ and verifying the signature of the state on every input
message.
When the functionality $\fattest$ is initialized, the initial state $\st_0$ is empty (or, $\epsilon$). % where $r$ denotes all the randomness that $g$ will ever use. 
Subsequent invocations of the functionality is done on input $w_\ctr$
% and program counter $\ctr$ 
using the $\compute$ command. 
The function $g$ is a deterministic mapping from $(\ctr,w_{\ctr},r_\ctr,\st_{\ctr-1})$ to
$(y_\ctr,\st_{\ctr})$, where $r_\ctr$ is the required randomness.
% picked by $\fattest$. 
The functionality picks randomness $r_\ctr$, evaluates $g$ and provide a signature on the function output $y_\ctr$ using the signing key $\sksign$.  
Furthermore, $(y_\ctr,\st_\ctr)$ is always given to party $P$ such that $\st_\ctr$ contains $r_\ctr$ in clear and 
%$g$ is a
%deterministic function mapping $(\ctr,w_{\ctr},r,\st_{\ctr-1})$ to
%$(y_\ctr,\st_{\ctr})$. 
%
%Note, that doing so 
this ensures that there is no information hidden from $P$ and we only assume correct execution of $g$.
That is, the ideal functionality can evaluate functions and provide signed outputs and these outputs could have anyway been computed by party $P$ given
knowledge of $g, w_\ctr, r_\ctr, \ctr, \st_\ctr$, which are all known to $P$. 
Thereby, we only assume that the functionality will sign the output of $g$ on the appropriate input and not hide any  data from $P$. This significantly weakens what is assumed from the trusted hardware.

%\vspace{0.3in}

%\aseem{Figure 8, shouldn't ideal functionality store g somewhere, or it is understood?}

\begin{tffbox}
\begin{mdframed}
\begin{center}
{\bf Functionality} $\fattest^{(\vk,\sksign)}$
\end{center}
%\vspace{.1in}
{\small

$\fattest$ interacts with a party $P$.

\begin{tiret}
       \item On input message $(\gcommit,g)$ from $P$,
% sample $r$ from $\zo^*$ at random, where $r$ is all the randomness that computation of $g$ will ever need.
       \begin{enumerate}
       \item Record $(\gcommit,\st_0)$, where $\st_0 = \epsilon$;
       \item Send $( \st_0, \token_g)$ to $P$, where $\token_g = \sign_\sksign(H(g))$.
       \item Ignore further $(\gcommit,g)$ messages.
       \end{enumerate}
	\item  On input message $(\compute,w_\ctr)$ from $P$, retrieve $\st_{\ctr-1}$, pick required randomness $r_\ctr$ and compute $g(\ctr,w_\ctr,r_\ctr,\st_{\ctr-1})$ to obtain $(y_\ctr,\st_{\ctr})$ such that $\st_\ctr$ contains $(y_\ctr, r_\ctr)$. Send $(y_\ctr,\ctr, \sign_{\sksign}(y_\ctr || \ctr),\st_\ctr)$ to $P$.
	       			
\end{tiret}
%\vspace{.1in}
} %SMALL
\end{mdframed}
\caption{\sl The Authentication functionality $\fattest^{(\vk,\sksign)}$.}
\label{fig:attestideal}
	%\vspace{-0.2cm}
\end{tffbox}

\subsection{Semi-honest security to malicious security}
\label{subsec:nextmessagefunction}

Our technique takes any semi-honest secure MPC protocol and converts
it into a malicious secure MPC protocol in the
$\fattest^{(\vk,\sksign)}-$hybrid model. The idea is to have messages
sent by every party $P_i$ to every other party $P_j$ in the
semi-honest protocol be computed by the corresponding
$\fattest^{(\vk_i,\sksign_i)}$ functionality interacting with $P_i$,
where $(\vk_i,\sksign_i)$ are keys used by the functionality.
These messages can be verified by functionality $\fattest^{(\vk_j,\sksign_j)}$ interacting with $P_j$. 
We assume that every $\fattest^{(\vk_i,\sksign_i)}$ knows the verification key $\vk_j$ used by functionalities of all other parties $P_j$ in a reliable manner. 
Later, we show how to achieve this through the use of remote attestation in the context of Intel SGX. We now set notation and describe the next message function of any semi-honest secure MPC protocol and how we modify it for our use.
\\\\
\noindent\textbf{Next message function.} Let $\pi(\cdot)$ be the next message function of any semi-honest secure MPC protocol. $\pi(\cdot)$ takes the following values as input - two party ids $i$ and $j$, input $x_i$, a round number $\ctr$, randomness $r_{i,j,\ctr}$ and $\transcript_i$, which includes the transcript of all messages sent and received by the party $P_i$ so far. 
Given these, $\pi(\cdot)$ outputs $y_{\ctr}^{i,j}$, which is the message that $P_i$ must send to $P_j$ in round $\ctr$ and also updates $\transcript_i$ appropriately. Additionally, $\pi(\cdot)$ takes message $y_{\ctr}^{j,i}$ sent by $P_j$ to $P_i$ at round $\ctr$ and update $\transcript_i$ with this message. We now describe how to modify $\pi(\cdot)$ to $\pi^*(\cdot)$ to incorporate checks to detect malicious behavior.
\\\\
\noindent\textbf{Modified next message function.} $\pi^*(\cdot)$, is the modified function that builds upon $\pi(\cdot)$ and we describe it for $P_i$.
%We will assume that every $\fattest^{g,(\vk_i,\sksign_i)}(P_i)$ knows and agrees upon $\pi(\cdot)$; we will show how to achieve this as well through attestation. 
%The modifications made to the next message function are now described:

\begin{enumerate}

\item For $\ctr = 1$, 
%$\pi^*(\cdot)$, picks an execution identity $\sid$. 
Let $x_i$ be the input of $P_i$ in $\pi(\cdot)$. Then,
%let $y^i_0$ denote $x_i$, and the randomness $r_i$ that will be used by $P_i$ in $\pi(\cdot)$. 
$(\ctr, x_i)$ is stored as $\st_1$ (also called as $\transcript^1_i$) and sent to $P_i$.
% after signing it with $\sksign_i$. 

\item When $\pi^*(\cdot)$ receives a message $M =
  (y_{\ctr}^{j,i},\ctr,\sigma)$ from party $P_j$, it runs
  $\verify_{\vk_j}((y_{\ctr}^{j,i},\ctr),\sigma)$. If verification
  succeeds, it appends $M$ to $\transcript_i$. Else, $P_i$ aborts.

\item $\pi^*(\cdot)$ on input $(\ctr, \st_{\ctr-1}, j)$ computes the
  next message from $P_i$ to $P_j$ as follows: It checks that
  $\st_{\ctr-1}$ contains a valid transcript of all messages computed
  so far. If it verifies, it picks randomness $r_{i,j,\ctr}$ and runs $\pi(\ctr, \st_{\ctr-1}, j, r_{i,j,\ctr})$ to compute next message $y^{i,j}_\ctr$ and updated state $\st_\ctr$ (containing $r_{i,j,\ctr}$). 
  Else it outputs $\bot$. 
Note that $\st_{\ctr-1}$ already contains
  input $x_i$, the input of party $P_i$.
\end{enumerate}~\\
\noindent\textbf{Malicious MPC in the $\fattest-$hybrid model.}
\label{sec:shtomcompiler}
The malicious \mpc protocol works as follows: Each party $P_i$ invokes $\fattest^{(\vk_i,\sksign_i)}$ with the command $\gcommit$ using function $\pi^*(\cdot)$ described above and sends the received token $\token^{(i)}_{\pi^*}$ to other parties $P_j$. It  receives similar tokens $\token^{(j)}_{\pi^*}$ from party $P_j$ and verifies it under $\vk_j$. Party $P_i$ aborts if any of these verifications fail. If all verifications succeed, it proceeds with running $\pi^*(\cdot)$ inside $\fattest^{(\vk_i,\sksign_i)}$ as described formally in Figure~\ref{fig:shtomprotocol}.

%whenever party $P_i$ must send a message according to $\pi(\cdot)$ to party $P_j$ in the underlying semi-honest secure protocol in round $\ctr$, $P_i$ invokes $\fattest^{\pi^*(\cdot)}$ along with the appropriate messages needed to compute the message $y_{\ctr}^{i,j}$. The functionality will check the validity of all messages and transcript and if correct, will produce $y_{\ctr}^{i,j}$ along with a signature on it (along with $\ctr,\sid$). This signed message is then passed on by $P_i$ to $P_j$ as the message in that round. 
%The complete protocol is in Figure \ref{fig:shtomprotocol}.

%\vspace{0.3in}

%%\aseem{(a) What is sid in Figure 9, is it just i? (b) In the fifth
%  step, it invokes F sub attest on (sid, Compute, j), is the last
%  argument j? Shouldn't it be the output of pi star? (c) The order
%  of ctr, sigma, y sub i,j is inconsistent, it won't typecheck :).}

\begin{tffbox}
\begin{mdframed}
\begin{center}
{\bf Protocol} $\protshtom$
\end{center}
%\vspace{.1in}
{\small

Party $P_i$ with input $x_i$ interacts with $\{P_j\}_{j\ne i}$ and $\fattest^{(\vk_i,\sksign_i)}$ and does the following:
\begin{tiret}
\item Invokes $\fattest^{(\vk_i,\sksign_i)}$ on $(\gcommit,\pi^*)$ to receive $( \st^{(i)}_0, \token^{(i)}_{\pi^*})$ and sends $\token^{(i)}_{\pi^*}$ to all parties $P_j$, $j \ne i$.

\item Receives $\token^{(j)}_{\pi^*}$ from $P_j$ and runs $\verify_{\vk_j}(H(\pi^*), \token^{(j)}_{\pi^*} )$ for all $j \in [n]\setminus i$. Aborts if one of these checks fail.

\item Invokes $\fattest^{(\vk_i,\sksign_i)}$ on $(\compute, x_i)$ to get $\transcript_i^1$ containing input $x_i$. %randomness $r_i$ that would be used to generate future messages from $P_i$.

\item When $P_i$ receives a message $M = (y_{\ctr}^{j,i},\ctr,\sigma)$ from party $P_j$, it invokes $\fattest^{(\vk_i,\sksign_i)}$ on  $( \compute, (y_{\ctr}^{j,i},\ctr,\sigma))$ and receives updated transcript or $\bot$ (and aborts).

\item When $P_i$ needs to send next message to $P_j$ it invokes $\fattest^{(\vk_i,\sksign_i)}$ on $( \compute, j)$ and receives $(y_{\ctr}^{i,j},\ctr, \sigma)$ along with updated transcript and randomness used. Here, $\sigma$ is a signature on $(y_{\ctr}^{i,j},\ctr)$ under $\sksign_i$. It sends $(y_{\ctr}^{i,j},\ctr, \sigma)$ to $P_j$.
   
   \item When $P_i$ has no more messages to send in $\pi(\cdot)$, it computes the output of the function from 
%transcript messages 
$\transcript_i$.
	       			
\end{tiret}
%\vspace{.1in}
} %SMALL
\end{mdframed}
\caption{\sl Malicious secure MPC $\protshtom$.}
\label{fig:shtomprotocol}
\end{tffbox}

%\vspace{0.3in}
\begin{tffbox}
\begin{mdframed}
\begin{center}
{\bf Functionality} $\fmpc^f(P_1,\cdots,P_n)$
\end{center}
%\vspace{.1in}
{\small

$\fmpc^f$ interacts with parties $\{P_1,\cdots,P_n\}$ and the adversary $\simu$.

\begin{tiret}
       \item On input message $x_i$ from $P_i$ record $x_i$ and ignore further $x_i$ from $P_i$
       \item Upon receiving $x_i$ from all $P_i, i\in [n]$, compute $y = f(x_1,\cdots,x_n)$ and send to $\simu$.
       \item Upon receiving $(i,\mathsf{Send})$ or $(i,\bot)$ from $\simu$, send $y$ or $\bot$ to $P_i$.

	       			\end{tiret}
%\vspace{.1in}
} %SMALL
\end{mdframed}
\caption{\sl The MPC functionality $\fmpc^f$.}
\label{fig:fideal}
%	\vspace{-0.5cm}
\end{tffbox}
%\vspace{-1cm}

\noindent\textbf{Malicious Security.} Next, we prove that if $\pi$ is secure against semi-honest adversaries, then the protocol described in Figure~\ref{fig:shtomprotocol} is an \mpc protocol secure against malicious adversaries with the same corruption threshold.  We prove the following result using the standard simulation paradigm in Appendix~\ref{app:proof}.

\begin{theorem}
\label{theorem:maliciousmpc}
Let $\pi(\cdot)$ be a semi-honest secure MPC protocol securely realizing $\fmpc^f$. Then, protocol $\protshtom$ described in Figure \ref{fig:shtomprotocol} securely realizes $\fmpc^f$ in the $\fattest^{(\vk_i,\sksign_i)}-$hybrid model (with $i \in [n]$) against malicious adversaries.
\end{theorem}

\subsection{Realizing $\fattest$}
\label{sec:fattest-sgx}
We note that the ideal functionality assumed out of the hardware can potentially be realized using various hardware platforms that provide code attestation and secure signing, e.g., STM32H7, MediaTek MT3620, CEC1702, ARMTrustZone, Intel SGX, etc. In this work, we provide an implementation of Aramis based on Intel SGX. 

SGX allows a host to create a protected region known as an enclave. Intel gives integrity guarantees, that is, the code and the data residing in the enclave, once attested, cannot be modified by the host or the operating system. 
When SGX receives a $\gcommit$ command (Figure~\ref{fig:attestideal}) for a function $g$, then it creates an enclave with code $g$.
Randomness $r_\ctr$ of Figure~\ref{fig:attestideal} can be sampled in SGX using {\tt sgx\_read\_rand} command. 
The attestation token $\token_g$ is generated by SGX communicating with Intel's Attestation Service (IAS) and this token is publicly verifiable given $g$ and  public verification key corresponding to Intel's Report Signing Key.
The key-pair $(\vk,\sksign)$ for ECDSA signature scheme is also generated inside the enclave and the verification key $\vk$ is sent as payload to IAS during the generation of the attestation token. 
The token $\token_g$ contains the verification key $\vk$ in the clear and this $\vk$ can be used to verify the signed outputs $y_\ctr$.
Now, on receiving the $\compute$ command, the enclave starts executing the code of $g$ and produces outputs signed under $\sksign$.

While running \mpc in the $\fattest$-hybrid, we require the enclave to reliably have verification keys used by enclaves of all other parties. This can be done by attaching the following prelude to $\pi^*$ (the code running inside SGX): Read the tokens of all parties, parse them to obtain the verification keys, and verify the signature on the tokens using verification key of Intel's Report Signing key. Note that since all the parties are running the same function $\pi^*$ (appended with this prelude), they can compute the hash of $\pi^*$ locally and compare it with the hash in the tokens (which has been signed by Intel's IAS) of all the other parties, proceeding only if they all match perfectly.

\subsection{Implementation challenges with Intel SGX}
\label{sec:challenges-aramis}

%\dg{SGX memory issues - [ReLU chunking, liveness ], ecall-ocall message passing payload optimizing, MAC}

We outline some of the key challenges in implementing \mpc between multiple SGX enclaves that involve multiple rounds of interaction and operate over large volumes of data.

\subsubsection{Memory constraints in SGX}
In SGX, all the enclave content, including code, and related data is stored in a special region of memory known as the Enclave Page Cache (EPC). The size of EPC is fixed in BIOS and can have a maximum size of 128MB. Typically, paging facilitates the execution of enclaves which cannot fit in EPC and any page that is evicted out is encrypted before storing it on unprotected memory \cite{intelsgxperf}. This additional overhead  has  detrimental effects on the overall performance of the enclave application. 
We reduce the working set of secure inference tasks to limit these overheads.
%To overcome this, we make changes to our code in the following way:
\begin{itemize}
	\item \textit{ReLU and MaxPool functions:} 
  % We split the computation of memory intensive non-linear functions into chunks  that fit  in EPC to avoid paging. For example, a secure ReLU computation that requires 120MB memory is split into 3 chunks  of 40MB each. Note that chunking increases the number of \mpc rounds and very small chunks are actually detrimental to performance.
  We split the computation of memory-intensive non-linear functions into chunks that fit in the EPC to avoid paging. However, lower chunk sizes  increase  the number of rounds, and so, the chunk sizes must be carefully selected. For \resnet, we  set the chunk sizes for ReLU and MaxPool layers to be 40 MB and 10 MB respectively. For our network configurations, the increase in rounds is justified by the elimination of paging costs and reduction in end-to-end runtimes.
	\item \textit{Convolution and Matrix Multiplication functions:} For the linear functions, we block the matrices into smaller ones, process the blocks, and aggregate them. We ensure that individual blocks fit in EPC. 
	\item \textit{Liveness Analysis:} Athos implements liveness analysis (Section~\ref{sec:athosopt}) which reduces the memory footprint of the compiled DNNs.  For example, the memory footprint of \resnet reduces from 1100 MB to 397 MB due to liveness analysis. When chunking and liveness analysis are done together, the memory footprint of \resnet comes down to 297MB.

\end{itemize}

%For the case of \resnet, we chose a chunksize of 40, 10 and 40 MiB for ReLU, MaxPool and Convolution, respectively. The number of rounds of a function increases linearly with the number of chunks needed to complete the function evaluation. For example, in \resnet, the ReLU with the maximum elements requires 5 chunks for 40 MiB each, making the rounds grow from 10 to 50. The values of chunksizes were chosen empirically to find a sweet-spot between increased number of rounds and decreased working set memory. In isolation to this chunking optimization, with liveness analysis, we observed that the memory footprint of \resnet comes down from 1100 MB to a mere 397 MB. Both these optimizations played a major role in drastically improving the performance of Aramis for our benchmarks.

\subsubsection{Porting Interactive Protocols to SGX}
To the best of our knowledge, we are the first work to implement highly interactive protocols in SGX and this comes with unique challenges. For example, whenever data is passed across the enclave's protected memory region, it has to be {\em marshalled} in/out of the region.
%\footnote{When pointers to memory are passed as parameters into the enclave via an $\mathsf{ecall}$, the referenced data block is {\em marshalled} into the enclave, specifically into the protected memory region that an enclave uses. Similarly, when a pointer to enclave data, residing in protected memory region, is passed outside an enclave via an $\mathsf{ocall}$, the referenced data block is marshalled out of the protected memory region.}. 
The performance of marshalling depends on the size of the parameters crossing the bridge. Larger parameters imply slower marshalling~\cite{intelsgxperf}, while smaller parameters increase the total numbers of cross-bridge calls (which have an overhead of their own). 
%For example, we observed that in order to send 1MB data in/out of the enclave, if 1048576 calls (OCALL-ECALL pair) are done with each carrying a payload of 1B, then it takes about 3.1 s for the calls alone, 16 calls with payload of 65KB each only take about 0.5 ms and 2 calls of 512KB each take 0.8 ms. Since we make calls of the order of 100000, 
Thus, we tune the payload size carefully. 
We also implement the techniques in~\cite{sealedglass}  for optimizing communication involving enclaves. 

\section{Experiments}\label{sec:experiments}

\noindent\textbf{Overview.} In this section, we present our experimental results. First, in Section \ref{subsec:bigbenchmarks}, we use \tool to securely compute inference on the ImageNet dataset using the following \tensorflow programs: \resnet\footnote{\url{https://github.com/tensorflow/models/tree/master/official/r1/resnet}} and \densenet\footnote{\url{https://github.com/pudae/tensorflow-densenet}}. 
%The computation is performed as a three-party secure computation protocol, where the model and query is secret shared amongst the three parties.
%We stress that no prior work has run MPC on networks of this scale. 
We also show that the performance of semi-honest and malicious protocols generated by \tool scale linearly with the depth of DNNs.
 Second, in Section \ref{sec:prior-comparison}, we show that \cryptflow\ outperforms prior works on secure inference of DNNs. 
%These experiments are on smaller DNNs that perform prediction over the MNIST and CIFAR datasets as prior work could only handle such small benchmarks.
 Next, we evaluate each component of \tool in more detail. 
In Section \ref{sec:athosexperiments}, we show that the fixed-point code generated by Athos matches the accuracy of floating-point  \resnet and \densenet.  We show in Section \ref{subsec:porthosexperiments} how the optimizations in Porthos help it outperform prior works in terms of communication complexity and overall execution time. In Section \ref{subsec:aramisexperiments}, we show the overhead of obtaining malicious secure MPC (over semi-honest security) using Aramis for GMW~\cite{gmw} and Porthos.  
We show Aramis-based malicious secure inference outperforms pure crypto-based  malicious secure protocols by huge margins in Section~\ref{sec:concurrent-comparison}.
Finally, in section \ref{subsec:realworldimpact}, we discuss two case-studies of running \cryptflow\ on DNNs for healthcare.
 We begin by providing details of the systems used to run our experiments.
\\\\
\noindent\textbf{System Details.} All our large benchmark experiments are in a LAN setting on 3.7GHz machines, each with 4 cores and with 16 GB of RAM running Linux Ubuntu 16.04. The measured bandwidth between each of the machines was at most 377 MBps and the latency was sub-millisecond. 
Since we wanted to use the same machines to benchmark both our semi-honest as well as our malicious secure protocols, we were constrained to use machines that had Intel SGX enabled on them - this led to machines that had considerably lower bandwidth between them (377 MBps) than those normally used by prior works in the area (e.g. \cite{aby3, quantizednn} used networks with bandwidth of 1.5 GBps). For Aramis, we used Intel SGX SDK version 2.4.
The compilation time of \tool is around 5 sec for \resnet, 35 sec for \densenet and 2 minutes for {{\textsc{ResNet200}}\xspace}.
% remains under 40 seconds for our benchmarks.
% \nc{A line here about the code being available...}

%Through this evaluation, we want to demonstrate the following:
%\begin{itemize}
%\item Section 6.1 shows that we can run state of the art neural networks securely using cryptflow.
%We show imagenet scale predictions using some of the most famous DNNs (ResNet50, VGG, MobileNets).
%No prior published work has run MPC on networks of this scale.
%\item Section 6.2 shows that cryptflow is far superior to other alternatives performance wise. 
%These experiments are on smaller MNIST/CIFAR DNNs as prior work has handled only those.
%Maybe we also show that cryptflow is much easier to use than other frameworks.
%\item Section 6.3 shows that Athos' scale exploration is critical for obtaining a good fixed-point network.
%We also show a sanity check that fixed-point runs much faster than floating-point in MPC.
%\item Section 6.4 uses microbenchmarks to evaluate Porthos against state-of-the-art 2PC and 3PC crypto protocols.
%\item Section 6.5 shows overhead of malicious security using SGX. 
%\end{itemize}
\subsection{Secure Inference on ImageNet}\label{subsec:bigbenchmarks}
%In this section, we show the power of \cryptflow, by demonstrating secure inference of \resnet and \densenet over the ImageNet dataset with over 1000 classes. 
We briefly describe our  benchmarks and then present performance results.
\begin{enumerate}
\item \resnet~\cite{resnet} is a network that follows the residual neural network architecture. 
The residual nodes employ ``skip connections'' or short cuts between layers.
% in order to avoid the problem of vanishing gradients while training.
 It consists of 53 convolution layers with filters of size up to $7\times 7$, and 1 fully connected layer of size $2048\times 1001$. The activation function between most layers is batch normalization (Appendix~\ref{appendix:batchnorm}) followed by ReLU. After the first convolutional layer, the activation function also includes a MaxPool.

\item \densenet~\cite{densenet} is a form of residual neural network that employs several parallel skips. It consists of 121 convolutional layers with filters of size up to $7\times 7$. The activation function between these layers is usually batch normalization, followed by ReLU. Some layers also use MaxPool or AvgPool. 

%\item \squeezenet~\cite{squeezenet} is a notoriously hard to train network and no prior work has considered evaluating this architecture on ImageNet with %fixed-point arithmetic.
%It consists of 26 convolutional layers with filters of size up to $3\times 3$. The activation function between these layers is usually ReLU and MaxPool. 
\end{enumerate}~\\
\noindent\textbf{Performance.} Table \ref{tab:bigbenchmarks} shows performance of \cryptflow\ on these benchmarks.
% within a range of 25--36 seconds with semi-honest security and 75--112 seconds with malicious security.
 We measure communication as total communication between all $3$ parties - each party roughly communicates a third of this value. The communication in semi-honest secure and malicious secure inference is almost the same. Thus demonstrating that ImageNet scale inference can be performed in about 30 seconds with semi-honest security and in under two minutes with malicious security. The malicious protocol of \tool is about 3x slower than the semi-honest version.

\begin{figure}
  \includegraphics[width=\linewidth]{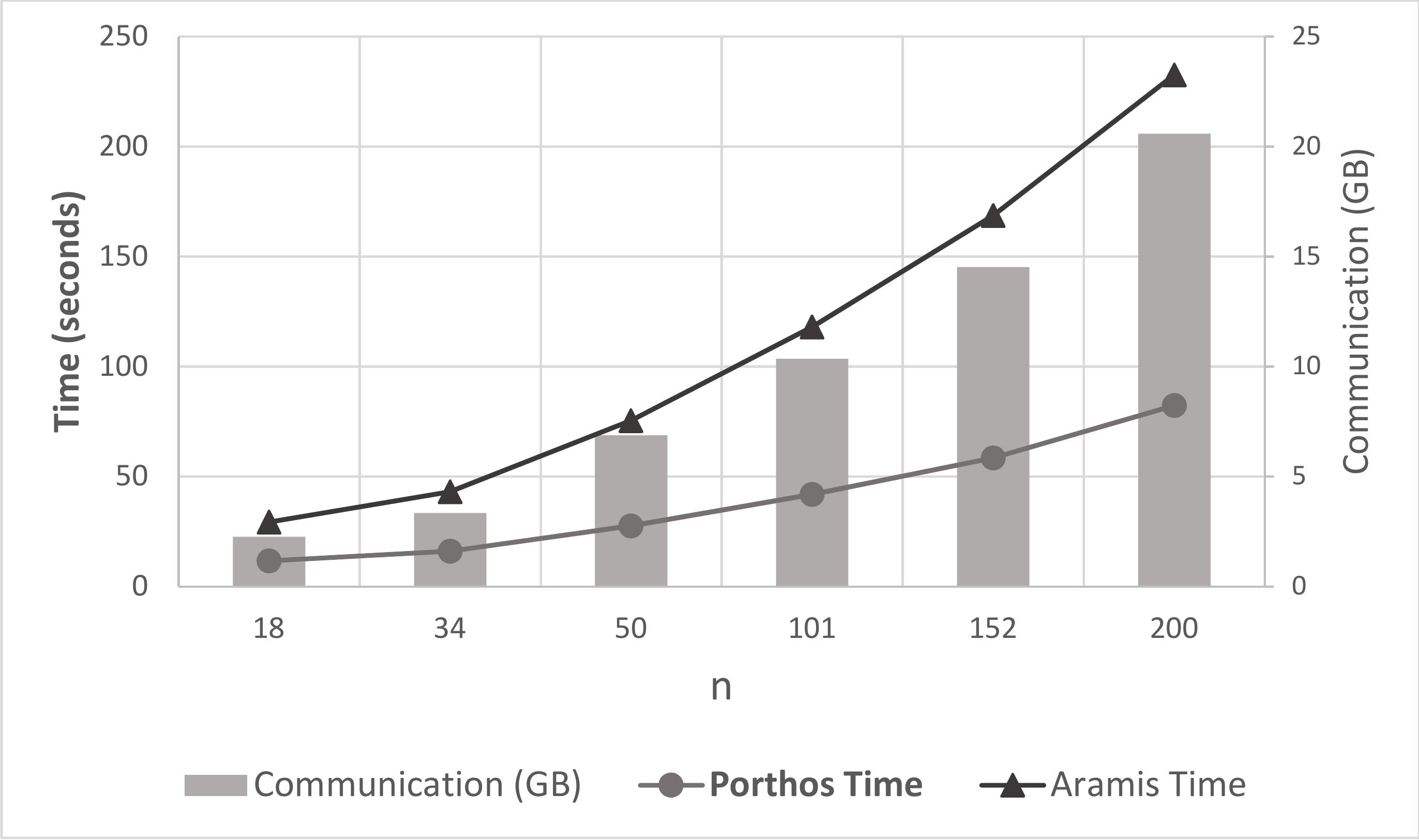}
	\caption{Scalability of {\cryptflow} on {\sc ResNet}-$n$.}
  \label{fig:scalingResnet}
\end{figure}

%\nc{We probably need to say something about QuantizedNN and differences}
\begin{table}
  \centering
%  \resizebox{0.7\columnwidth}{!}{

      \begin{tabular}{|c|c|c|c|c|}
    \hline
    Benchmark & Semi-Honest (s) &  Malicious (s) & Comm. (GB)  \\
    \hline
    $\resnet$ & $25.9$ & 75.4 &$6.9$\\ \hline
    $\densenet$ & $36.0$ & 112.9 &$10.5$\\ 	\hline
%    $\squeezenet$ & $11.3$ & 29.0 &$2.6$\\ 	\hline

\end{tabular}
%}
 \caption{\cryptflow: ImageNet scale benchmarks.}
\label{tab:bigbenchmarks}
%\tableup
	%\vspace{-0.5cm}
\end{table}~\\
\noindent\textbf{Scalability.} We show that the running time of \tool-based protocols increases linearly with the depth of DNNs. We compile \textsc{ResNet-$n$} (where $n$, the approximate number of convolutional layers, varies from 18 to 200) with \tool and evaluate with both semi-honest (Porthos) and malicious secure protocols (Aramis) in Figure \ref{fig:scalingResnet}. Our largest benchmark here is \textsc{ResNet-$200$}, the deepest version of \textsc{ResNet} on the ImageNet dataset \cite{he2016identity}, which has 65 million parameters. Other \textsc{ResNet}-$n$ benchmarks have between 11 to 60 million parameters \footnote{Specifically, 11, 22, 25, 44 and 60 million parameters for \textsc{ResNet}-$n$ for $n=$ 18, 34, 50, 101, and 152 respectively.}. We observe that the communication and runtime increase linearly with depth. Even with increasing depth, the overhead of malicious security (over semi-honest security) remains constant at about 3X. 

\subsection{Comparison with prior work}
\label{sec:prior-comparison}
In this section, we show that \tool outperforms prior works on secure inference of DNNs
 on the benchmarks they consider, i.e., tiny 2--4 layer DNNs over the MNIST and CIFAR-10 datasets. We stress that these benchmarks are very small compared to the ImageNet scale DNNs discussed above. % and hence may not be an accurate reflection of the power of \cryptflow. 
In order to provide a fair comparison, for these experiments, we use a network with similar bandwidth as prior works (1.5 GBps) and machines with similar compute (2.7 GHz). 
%For protocols whose code is publicly available (e.g. SecureNN~\cite{securenncode}), we ran the protocol on our benchmark machines and the times reported are those; for others, the times are from the respective papers that use similar machines. 

Table~\ref{tab:porthosvspriormnist} shows that Porthos outperforms prior (ABY$^3$ \cite{aby3}, Chameleon\footnote{Chameleon is a 2PC protocol in the online phase but requires a trusted third party in the offline phase. We report overall time here.}~\cite{chameleon}, and SecureNN \cite{securenn}) and concurrent (QuantizedNN~\cite{quantizednn}) semi-honest secure 3PC works on the MNIST dataset. 
It is well-known that 3PC-based techniques like Porthos are much faster than techniques based on 2PC and FHE. We relegate comparison between Porthos and 2PC/FHE
works to Appendix~\ref{unfaircomp}. We omit comparisons with~\cite{trident,astraccsw} as their published MSB protocol was incorrect~\cite{astraeprint}.
%
%Table \ref{tab:porthosvspriorcifar10} compares Porthos with prior 2PC work.  We omit some prior works (e.g., \cite{secureml,hycc,chameleon}, etc.) in these tables as they are slower than %Gazelle and do not provide additional insights. 
%We note that other than ABY$^3$ \cite{aby3}, QuantizedNN \cite{quantizednn}, and SecureNN \cite{securenn}, other works are for 2-party secure computation. 
%As can be seen from the tables, the 2PC systems are much slower than the 3PC systems and Porthos performs better than other 3PC systems. For instance, for a 4-layer CNN for MNIST %from MiniONN~\cite{minionn}, the best 2PC-backend Gazelle takes 810 ms, prior best 3PC work takes 47 ms, and Porthos takes 34 ms. 
%Further, some of these benchmarks employ much higher bandwidth than our system  -- e.g. ABY$^3$~\cite{aby3} and QuantizedNN~\cite{quantizednn} make use of a network with %1.5GBps bandwidth which is {\em more than 3 times faster than our network}; however the code of these works are not public and hence we could not reproduce their results on our %benchmark machines. Even then, our performance numbers either beat prior times or are only marginally slower that accounting for bandwidth difference would make our protocol faster. 

\begin{table}
  \centering
%  \resizebox{\columnwidth}{!}{

      \begin{tabular}{|c|c|c|c|c|c|}
    \hline
    Benchmark & \cite{aby3}  & \cite{quantizednn} & \cite{securenn} & \cite{chameleon} & Porthos \\
    \hline
	Logistic Regression & $4$ & $-$ & $-$ & -  & $2.7$\\
    \hline
	SecureML (3 layer) & $8$ & $20$ & $17$ & - & $8$\\
	\hline
	MiniONN (4 layer)  & -  & $80$ & $47$ & 2240 & $34$\\
	\hline
   LeNet (4 layer) & - & $120$ & $79$ & - & $58$ \\
	\hline
\end{tabular}
%}
 \caption{Comparison with 3PC on MNIST dataset with ABY$^3$ \cite{aby3}, QuantizedNN~\cite{quantizednn}, SecureNN \cite{securenn}, and Chameleon~\cite{chameleon}. All times in milliseconds.}
\label{tab:porthosvspriormnist}
%\tableup
	%\vspace{-0.5cm}
\end{table}

%To put Aramis in comparison with concurrent actively secure inference works, we compare Aramis with QuantizedNN \cite{quantizednn} in table \ref{tab:aramisvsspdz} on the 4 networks described in SecureNN \cite{securenn}. We used MP-SPDZ framework \cite{mpspdz} to run QuantizedNN on these 4 benchmarks with malicious security in an honest majority setting. MP-SPDZ repository already has these 4 benchmarks written for QuantizedNN which we used for this comparison. However, it only implements 8-bit quantization with networks stripped out of all ReLU layers, while for Aramis we ran full networks with 64-bit quantization. In doing so we are being unfair to Aramis and the execution time of MP-SPDZ with ReLUs would be much worse
%than what is shown here.

\begin{comment}
\begin{table}
  \centering
  \resizebox{0.7\columnwidth}{!}{

      \begin{tabular}{|l|c|c|c|c|}
    \hline
    Benchmark & Aramis Gains over Mal. QuantizedNN\\
    \hline
	SecureNN N/W A & $9$x \\
	\hline
	SecureNN N/W B & $32.7$x \\
	\hline
	SecureNN N/W C & $34.1$x \\
	\hline
	SecureNN N/W D & $7.9$x \\
	\hline

\end{tabular}
}
 \caption{MNIST dataset -- Aramis vs Malicious QuantizedNN}
\label{tab:aramisvsspdz}
%\tableup
	%\vspace{-0.5cm}
\end{table}
\end{comment}

%these benchmarks are at a regime where computation is not an issue (as the matrix multiplications performed by parties are small) and are only bandwidth constrained. This is in contrast to larger benchmarks where both compute and communication are overheads. 
\subsection{Athos experiments}
\label{sec:athosexperiments}
\noindent\textbf{Accuracy of Float-to-Fixed.} 
We show that Athos generated fixed-point code matches the accuracy of floating-code on \resnet and \densenet in Table~\ref{tab:fixed-accuracy}.
The table also shows the precision or the scale that is selected by Athos (Section~\ref{subsec:athosquantizer}). We observe that different benchmarks require different precision to maximize the classification accuracy and that the technique of ``sweeping'' through various precision levels is effective.
We show how accuracy varies with precision in Appendix~\ref{appendix:accuracies}.
Evaluating accuracy also helps validate the correctness of our compilation~\cite{frigate}. 

\begin{table}
\centering
% \resizebox{0.7\columnwidth}{!}{
\begin{tabular}{|c|c|c|c|c|c|}
\hline
Benchmark & Float & Fixed & Float & Fixed & Scale \\
          & Top 1 & Top 1 & Top 5 & Top 5 & \\
\hline
$\resnet$     & 76.47 & 76.45 & 93.21 & 93.23 & 12 \\ \hline
$\densenet$   & 74.25 & 74.33 & 91.88 & 91.90 & 11 \\ \hline
%$\squeezenet$ & 55.86 & 55.92 & 79.18 & 79.24 & 10 \\ \hline
\end{tabular}
%}
\caption{Accuracy of fixed-point vs floating-point.}
\label{tab:fixed-accuracy}
%\tableup
%\vspace{-0.8cm}
\end{table}~\\
\noindent\textbf{Modularity.} 
Since \tool is modular, we can compile it to various \mpc backends. To demonstrate this ability, we also add a 2PC semi-honest secure protocol ABY~\cite{aby} to \tool. 
The performance with this backend is in Table \ref{tab:2pcnumbers}.
We ran logistic regression (LR) as well as a small LeNet network~\cite{lenet} which comprises of 2 convolutional layers (with maximum filter size of $5\times 5$) and 2 fully connected layers, with ReLU and MaxPool as the activation functions.
% We see that performance of 2PC is much worse than 3PC Porthos.. 
 This evaluation shows that \cryptflow\ can be easily used for a variety of backends. 
 % - however the current state-of-the-art performance of 2PC makes it difficult to execute the large DNNs described in Section \ref{subsec:bigbenchmarks}. 
%One could potentially implement the functions in Table~\ref{tab:smf} with
% state-of-the-art 2PC backends such as Delphi~\cite{delphi}. The code of Delphi is not publicly available and hence we could not do the same.

\begin{table}
  \centering
%  \resizebox{0.7\columnwidth}{!}{

      \begin{tabular}{|c|c|c|}
    \hline
    Benchmark & \cryptflow\ (s) & Communication (MB) \\
    \hline
	$\mathsf{Logistic Regression}$ & $0.227$ & $25.5$\\
	\hline
    $\mathsf{LeNet}$ $\mathsf{Small}$ & $47.4$ & $2939$ \\ 	\hline

\end{tabular}
%}
 \caption{\cryptflow\ compilation to 2PC on MNIST.}
\label{tab:2pcnumbers}
%\tableup
%	\vspace{-0.8cm}
\end{table}

%A few other interesting observations follow. After a certain threshold in scaling factor, accuracies drop dramatically. This is expected as beyond this level of precision, we lose information %due to the overflow of underlying values when performing computation such as matrix multiplication. Interestingly, accuracy doesn't always improve with larger bits of precision even until %this threshold - e.g., the best Top 1 accuracy on the \resnet\ network is obtained at 12 bits of precision.

\subsection{Porthos experiments}\label{subsec:porthosexperiments}
Since Porthos builds on SecureNN, we compare them in mode detail.
As described earlier, Porthos improves over the communication complexity of SecureNN~\cite{securenn} both for convolutional layers as well as for non-linear activation functions.
We have already compared SecureNN and Porthos on benchmarks considered in SecureNN in Table~\ref{tab:porthosvspriormnist}.
Additionally, we also compare Porthos and SecureNN on ImageNet scale benchmarks in Table \ref{tab:porthosvssecurenn}. For this purpose, we add the code of SecureNN available at~\cite{securenncode} as another backend to \tool. These results show that Porthos improves upon the communication of SecureNN by a factor of roughly 1.2X--1.5X and the runtime by a factor of roughly 1.4X--1.5X.
% on these benchmarks.

\begin{table}
  \centering
  \resizebox{\columnwidth}{!}{

      \begin{tabular}{|c|c|c|c|c|}
    \hline
    Benchmark & SecureNN & Porthos& SecureNN & Porthos \\
     & (s)  & (s) & Comm. (GB) & Comm. (GB) \\
	\hline
	$\resnet$ & $38.36$ & $25.87$& $8.54$& $6.87$ \\
	\hline
    $\densenet$ & $53.99$ & $36.00$& $13.53$ & $10.54$ \\ 	\hline
  %  $\squeezenet$ & $16.55$ & $11.28$& $3.88$ & $2.63$ \\ 	\hline

\end{tabular}
}
 \caption{Porthos vs SecureNN.}
\label{tab:porthosvssecurenn}
%	\vspace{-0.5cm}
%\tableup
\end{table}

\subsection{Aramis experiments}\label{subsec:aramisexperiments}
We applied Aramis to both the 2-party GMW protocol~\cite{gmw} (using the codebase~\cite{gmwcode}, based on~\cite{gmwpaper}) as well as Porthos. 
The results for different functions using the GMW protocol are presented in Table \ref{tab:gmwport}. $\mathsf{IP}_n$ denotes the inner product of two $n$-element vectors over $\bbF_2$, $\mathsf{Add}_{32}$ and $\mathsf{Mult}_{32}$ denote addition and multiplication over $32$ bits respectively, and $\mathsf{Millionaire}_{32}$ denotes the millionaires problem that compares two $32-$bit integers $x$ and $y$ and outputs a single bit denoting whether $x>y$.  The overheads of Aramis-based malicious security, are within $54\%$ of the semi-honest protocol. Table~\ref{tab:bigbenchmarks} and Figure~\ref{fig:scalingResnet} evaluate Aramis with Porthos.
%Aramis in this table denotes the semi-honest GMW protocol ported into Intel SGX to provide malicious security. Evaluation of Aramis applied to Porthos 

% As can be seen, all overheads in this case, are within $54\%$ of the semi-honest protocol. The evaluation of Aramis on

%Since these benchmarks are small, all of the code and data fit inside the SGX enclave without any requirement for paging and hence overheads are also minimal. For ImageNet scale %benchmarks, the overheads are higher (Table~\ref{tab:bigbenchmarks}).
%Since these benchmarks are much larger, we have to deal with well-known paging issues that occur when using Intel SGX with large data~\cite{intelsgxperf}; here we incur overheads of %about 3X over semi-honest protocols. 

\subsubsection{Comparison with crypto-only malicious \mpc}
\label{sec:concurrent-comparison}
We demonstrate that Aramis based malicious secure protocols are better suited for large scale inference tasks compared to pure cryptographic solutions. 
We compare the performance of Porthos compiled with Aramis and the concurrent work of QuantizedNN~\cite{quantizednn} that uses the MP-SPDZ~\cite{mpspdz} framework to also provide a malicious secure variant of their protocol. Both these approaches provide security for the same  setting of 3PC with 1 corruption. On the four MNIST inference benchmarks A/B/C/D in the MP-SPDZ repository, Aramis is 10X/46X/44X/15X faster.
% Furthermore, in these performance measurements, we are being unfair to Aramis: MP-SPDZ strips out all the ReLUs from its performance estimates and its %performance would be much worse in their presence. However, the Aramis time measurements include the time to compute ReLUs, which can be 80\% of the %total execution time. If one were to evaluate ReLUs with MP-SPDZ then the speedups of Aramis would be even higher.
% \nc{Should we clarify somewhere that earlier mentions of QuantizedNN were for semihonest security and here alone we are referring to their malicious protocol? To prevent against a reviewer thinking that all are malicious protocols?}

\begin{table}
  \centering
 % \resizebox{0.7\columnwidth}{!}{

      \begin{tabular}{|c|c|c|c|}
    \hline
    Benchmark & GMW (ms) & Aramis (ms) & Overhead \\
    \hline
	$\mathsf{IP}_{10,000}$ & $464$ & $638$ & $1.37$x\\
	\hline
    $\mathsf{IP}_{100,000}$ & $2145$ & $3318$ & $1.54$x \\ 	\hline
    $\mathsf{Add}_{32}$ & $279$ & $351$ & $1.25$x \\ 	\hline
    $\mathsf{Mult}_{32}$ & $354$ & $461$ & $1.30$x \\ 	\hline
    $\mathsf{Millionaire}_{32}$ & $292$ & $374$ & $1.28$x \\ 
	\hline

\end{tabular}
%}
 \caption{Semi-honest GMW vs Malicious Aramis.}
\label{tab:gmwport}
	%\vspace{-0.5cm}

%\tableup
\end{table}

\subsection{Real world impact}\label{subsec:realworldimpact}
%Our evaluation thus far has been based on ML models and datasets in image recognition. In this section, 
We discuss our experience with using {\cryptflow} to compile and run DNNs used in healthcare. These DNNs are available as pre-trained Keras models. We converted them into \tensorflow using~\cite{kttf} and compiled the automatically generated \tensorflow code with \tool.

\paragraph{Chest X-Ray} In \cite{chestxray2018}, the authors train a {\densenet} to predict lung diseases from chest X-ray images. They use the publicly available NIH dataset of chest X-ray images and end up achieving an average AUROC score of 0.845 across 14 possible disease labels.
% We took their publicly available pretrained keras model, converted it to tensorflow~\cite{kttf}, and compiled the \tensorflow code with {\cryptflow}.
During secure inference, we observed no loss in accuracy and the runtime is similar to the runtime of \densenet for ImageNet.
%Both the latency and accuracy of this inference is similar to the one we report for {\densenet} and . We also end up achieving a similar AUROC score.
%\nishant{Todo on me to find this AUROC score for compiled code and double check that the performance is indeed identical to \densenet}

\paragraph{Diabetic Retinopathy CNN} Diabetic Retinopathy (DR), one of the major causes of blindness, is a medical condition that leads to damage of retina due to diabetes \cite{janakirammsv2017}. In recent times, major tech companies have taken an interest in using DNNs for diagnosing DR from retinal images \cite{janakirammsv2017,googleDRPaper}. 
%We used one such DNN pretrained in keras, converted the same to \tensorflow~\cite{kttf}, and then ran {\cryptflow} on it to run it securely using Porthos as our backend. 
Predicting whether a retina image has DR or not can be done securely in about 30 seconds with \tool.
%\nishant{Add more on this: performance, accuracy metric? but we only had 100 images, however google paper has auroc score.}

\section{Related Work}\label{sec:related}

\noindent\textbf{High level languages.} \tool is the first system to compile pre-defined \tensorflow code to secure \mpc protocols. There have been prior works that compile from lower-level, domain-specific languages to \mpc. Examples include Fairplay~\cite{fairplay}, Wysteria~\cite{wysteria}, ObliVM~\cite{oblivm}, CBMC-GC~\cite{cbmcgc}, SMCL~\cite{smcl},~\cite{lambdaps}, Sharemind~\cite{sharemind}, EzPC~\cite{ezpc}, and SPDZ~\cite{spdzcompiler}. Reimplementing large DNNs in the input format of these tools is a formidable task. PySyft~\cite{pysyft} and TF-Encrypted~\cite{tfe} are ongoing efforts that also aim to compile DNNs to \mpc protocols. 
%easier to write DNNs for secure ML tasks while providing modularity to plug in various MPC backends. 
In contrast to \tool that compiles standard \tensorflow code, these works require reimplementing the DNNs in a dialect of PyTorch/\tensorflow.
To the best of our knowledge, these systems have not been evaluated on ImageNet scale tasks.
%In addition, they don't support various compiler-level optimizations 
% like ReLU Maxpool switching 
%that \tool does and haven't been evaluated over large neural networks.
%In particular, the float-to-fixed compilation must be done manually by a user of these frameworks.
%Moreover, Tensorflow hides a lot of low-level details from the users. Hence, a user needs to be aware of all vagaries of Tensorflow while performing a faithful translation.
%Most of these frameworks compile a high level language into exclusively either an arithmetic or a boolean circuit, which is then securely computed using one of the numerous protocols~\cite{yao,gmw}. This is known to provide poor performance even for small real-world benchmarks that mix arithmetic and boolean computation, let alone something as complex as Tensorflow. While EzPC~\cite{ezpc}, SPDZ~\cite{spdzcompiler}, and HyCC~\cite{hycc} provide support for securely computing a mix of both arithmetic and boolean functions (e.g. using the ABY protocol~\cite{aby}). 
% Thus, prior systems fall short of providing the type of support required to run realistic ML benchmarks without any modifications to their original code. 
\\\\
\noindent\textbf{Fixed-point in \mpc.}
% Athos is the first float-to-fixed converter that has been designed for secure inference.
Although the use of fixed-point for secure computations is well-known~\cite{valeriaRidge},
prior works on secure inference have addressed the float-to-fixed problem by either
generating a fixed-point model by hand~(\cite{secureml,minionn,gazelle,aby3,securenn,delphi,chameleon}),
or by using non-standard training algorithms that output fixed-point models~(\cite{xonn,nitin}).
Both of these approaches are unsatisfactory. In particular, 
some of the challenges that one would face with the latter include: a) the need to train again on the whole training data which is both computationally expensive, and impossible if the training data is unavailable; and
b)
% the introduction of new nodes that deal with integers changes the network and can cause the training procedure 
%(a non-convex optimization problem) to diverge.
training algorithms that generate integer models  is still an active research area and an overwhelming majority of ML training algorithms still generate floating-point models. 
Athos alleviates all these problems by working with a trained model and being completely oblivious to the training procedure.
%Athos takes as input a pretrained floating-point model which can be obtained by standard training algorithms.
%Hence, we do not need to run training again (which can potentially take days) or modify the training algorithms. 
The ML users can train their networks in the manner they see fit and then use Athos to get  fixed-point code.
Finally, even with retraining, \tensorflow-generated binary/integer networks suffer significant accuracy loses~\cite{tflite} whereas
Athos matches the accuracy of floating-point models.
\\\\
\noindent\textbf{Float-to-fixed.}
The research literature in float-to-fixed for digital signal processors is rich and spans several decades.
However, it is only recently that these schemes have been adapted to machine learning.
%For example, SeeDot~\cite{seedot} is a strongly typed intermediate language which can express ML models and can be compiled to fixed-point code. 
%However, the SeeDot compiler assigns different precision to different parameters of a ML model in the generated fixed-point code.
%Hence, the generated program leaks private information about the model parameters and is unsuitable for secure machine learning.
Some recent float-to-fixed schemes~\cite{seedot,qualcom,tflite} show promise by quantizing floating-point models to 8-bit or 16-bit integers. One could potentially use one of these systems in place of our float-to-fixed component -- however, their compatibility with \mpc protocols~\cite{aby3,securenn} is unclear. Additionally, since we use higher bit-width of 64, not surprisingly, the accuracy of \tool is better.  
%Another approach is to use \tensorflow's ``post-training-quantization"\footnote{\url{https://www.tensorflow.org/lite/performance/post_training_quantization}} support that converts a %floating-point model
%to a model over 8-bit and 32-bit integers. However, at inference time, to preserve accuracy, all operations are still performed in floating-point arithmetic, which are very slow in \mpc %(Table~\ref{tab:floatvsfixed}).
%Similarly, the quantized  \squeezenet~\cite{squeezenet} models are stored as integers but are converted to floating-point at inference time.
% and lose efficiency.
%\nc{The above two paragraphs could potentially be condensed into one.} 
\\\\
\noindent\textbf{Secure Machine Learning.} There has been a flurry of recent results (\cite{sml1,codedprivateml,sml3,sml4,chiron}) in the area of secure machine learning, both in the 2-party~\cite{shafindss,valeriamatrix,cryptonets,minionn,gazelle,delphi,helen}, as well as in the 3-party setting~\cite{chameleon,aby3,securenn,quantizednn}. The most relevant to our work are ABY$^3$~\cite{aby3} and SecureNN~\cite{securenn} that both provide 3-party semi-honest secure computation protocols for a variety of neural network inference and training algorithms, with somewhat similar performance guarantees. Porthos, our 3-party semi-honest protocol, outperforms both these works. We also remark that there have been other recent works~\cite{xonn,codedprivateml,nitin,outsourcingprivateml,chet,nhe,nhe2,leviosa}, that modify the inference or training algorithms in order to obtain performance benefits.  These are applicable only to specialized benchmarks. For example, the works that use fully homomorphic encryption (e.g.,~\cite{chet,nhe,nhe2})  do not support secure evaluation of ReLUs, XONN~\cite{xonn} requires DNNs to have binary weights, etc.
 On the other hand, we focus on standard inference algorithms and \tool has much wider applicability.
\\\\
\noindent\textbf{Hardware-based security.}  Our work is the first to provide experimentally validated malicious secure inference of ML algorithms at the scale of \resnet. As discussed earlier, we achieve this by relying on minimally secure hardware to provide integrity. Prior works that use hardware enclaves for secure computation~\cite{vc3, obliviousmpml, GuptaFC16, BahmaniFC17,gcsgx,ndss1,ndss2,ndss3,slalom,opaque,chiron} assume that the enclave hides all data residing in it from the host. Thus, unlike Aramis, these systems are not secure against an adversary that can observe the SGX state. 
 The only prior work that assumes a weaker trust assumption from the hardware is that of~\cite{sealedglass}. Similar to our work, they assume that the hardware provides integrity. However, their work is  in the context of zero-knowledge proofs and other fundamentally asymmetric primitives that require only one enclave
and not interactive protocols between multiple enclaves. %ABY$^3$~\cite{aby3} provides a (purely cryptographic) method to convert their semi-honest secure protocol to a maliciously secure protocol. However, their method is both specialized to their protocol and is also not tested for performance.

\section{Conclusion}\label{sec:conclusion}
\cryptflow\ is the first end-to-end system that translates high-level
\tensorflow inference code to MPC protocols. It has 3 components - a) compiler from \tensorflow to \mpc, b) an improved semi-honest 3PC protocol for DNNs, and c) a generic technique to convert semi-honest secure protocols to malicious secure ones.
% that makes a minimal trust assumption in hardware. 
Using \cryptflow, we demonstrate the first instance of secure inference on large benchmarks such as \resnet\ and \densenet\ on the ImageNet dataset
with both semi-honest (in about thirty seconds) and malicious security (in less than two minutes).
 \cryptflow's modular design supports a variety of backends, and we hope that it can serve as a testbed for benchmarking new MPC protocols in the area. 

Going forward, we would like to plugin protocols like SPDZ~\cite{mpspdz} and Delphi~\cite{delphi} in \cryptflow.
%hope that protocol developers like~\cite{aby3,delphi} would integrate their work as backends to \cryptflow.
Our more ambitious goal is to extend \cryptflow\ to support
\tensorflow training.
%% We would like to consider compiling \tensorflow training code as
%% well.
It is a challenging problem since in the absence of the GPU support, the
overheads of MPC protocols for secure training can be prohibitive.
%% However, none of the existing  \mpc protocols~\cite{secureml,securenn,aby3} for training have GPU support and thus the overheads of secure training are huge.

\section{Acknowledgements}\label{sec:ack}
We thank our shepherd Xiao Wang, and anonymous reviewers for their valuable feedback.
We also thank  Sridhar Gopinath, Aayan Kumar,  Wonyeol Lee,
Sundararajan Renganathan, and Kapil Vaswani for helpful discussions.

%
% The next two lines define the bibliography style to be used, and the bibliography file.
\bibliographystyle{IEEEtranS}
\bibliography{main}

%\newpage
% 
% If your work has an appendix, this is the place to put it.
\appendix
\subsection{Algorithms used by Porthos}\label{appendix:porthos}
The additional algorithms that reshape filters, input, and output, used by Porthos are shown in Algorithms \ref{algo:reshapefilter}, \ref{algo:reshapeinput}, and \ref{algo:reshapeoutput}.

\begin{algorithm}[h]
\KwIn{$X \in \bbZ_L^{f \times f}$.}
\KwOut{$Z \in \bbZ_L^{f^2 \times 1}$.}
1. \textbf{for} $i = \{0, ..., f-1\}$ do \\
2. \hspace{10mm} \textbf{for} $j = \{0, ..., f-1\}$ do \\
3. \hspace{10mm} \hspace{10mm} $Z[i \cdot f + j] = X[i][j]$\\
4. \hspace{10mm} \textbf{end for}\\
5. \textbf{end for}

    \caption{{ ReshapeFilter} \label{algo:reshapefilter}}

\end{algorithm}

\begin{algorithm}[h]
\KwIn{$X \in \mathbb{Z}_L^{m \times m}$.}
\KwOut{$Z \in \mathbb{Z}_L^{n^2 \times f^2}$ where $n = m-f+1$.}
\textbf{Global Information}: Filter dimension $f$.\\
1. \textbf{for} $i = \{0, ..., m-f\}$ do \\
2. \hspace{10mm} \textbf{for} $j = \{0, ..., m-f\}$ do \\
3. \hspace{10mm} \hspace{10mm} \textbf{for} $k = \{0, ..., f-1\}$ do \\
4. \hspace{10mm} \hspace{10mm} \hspace{10mm} \textbf{for} $l = \{0, ..., f-1\}$ do \\
5. \hspace{10mm} \hspace{10mm} \hspace{10mm} \hspace{10mm} $Z[i \cdot (m-f+1) + j][k\cdot f+j] = X[k+i][l+j]$\\
6. \hspace{10mm} \hspace{10mm} \hspace{10mm} \textbf{end for}\\
7. \hspace{10mm} \hspace{10mm} \textbf{end for}\\
8. \hspace{10mm} \textbf{end for}\\
9. \textbf{end for}

    \caption{{ ReshapeInput} \label{algo:reshapeinput}}

\end{algorithm}

\begin{algorithm}[h]
\KwIn{$X \in \bbZ_L^{n^2 \times 1}$.}
\KwOut{$Z \in \bbZ_L^{n \times n}$.}
1. \textbf{for} $i = \{0, ..., n-1\}$ do \\
2. \hspace{10mm} \textbf{for} $j = \{0, ..., n-1\}$ do \\
3. \hspace{10mm} \hspace{10mm} $Z[i][j] = X[i\cdot n + j]$\\
4. \hspace{10mm} \textbf{end for}\\
5. \textbf{end for}

    \caption{{ ReshapeOutput} \label{algo:reshapeoutput}}

\end{algorithm}

\subsection{Batch Normalization}\label{appendix:batchnorm}
Batch Normalization \cite{batchNormPaper} is used to normalize the inputs to intermediate layers across a mini-batch of images. 
For a batch $B$ of inputs, let $\mu_B$ and $\sigma_B^2$ be the mean and the variance respectively. 
For an input $x$, the output of the batch normalization layer is defined as
\[ BN(x) = \gamma\frac{\left( x - \mu_B \right)}{\sqrt{\sigma_B^2 + \epsilon}} + \beta \]
where $\gamma$ and $\beta$ are the model parameters learned during training phase. In the inference phase, $\mu_B$ and $\sigma_B^2$ represent the mean and variance of the entire training dataset.

%\pagebreak
\subsection{Accuracy of Athos}\label{appendix:accuracies}

%\pagebreak
%\begin{comment}
\section{Accuracy of Athos}\label{appendix:accuracies}

% This section presents Top 1 and Top 5 accuracies of Athos on \densenet\ running on ImageNet.
In this section, we present the Top 1 and Top 5 accuracies of Athos on the ImageNet dataset. 

\begin{figure}
  \includegraphics[width=0.8\linewidth]{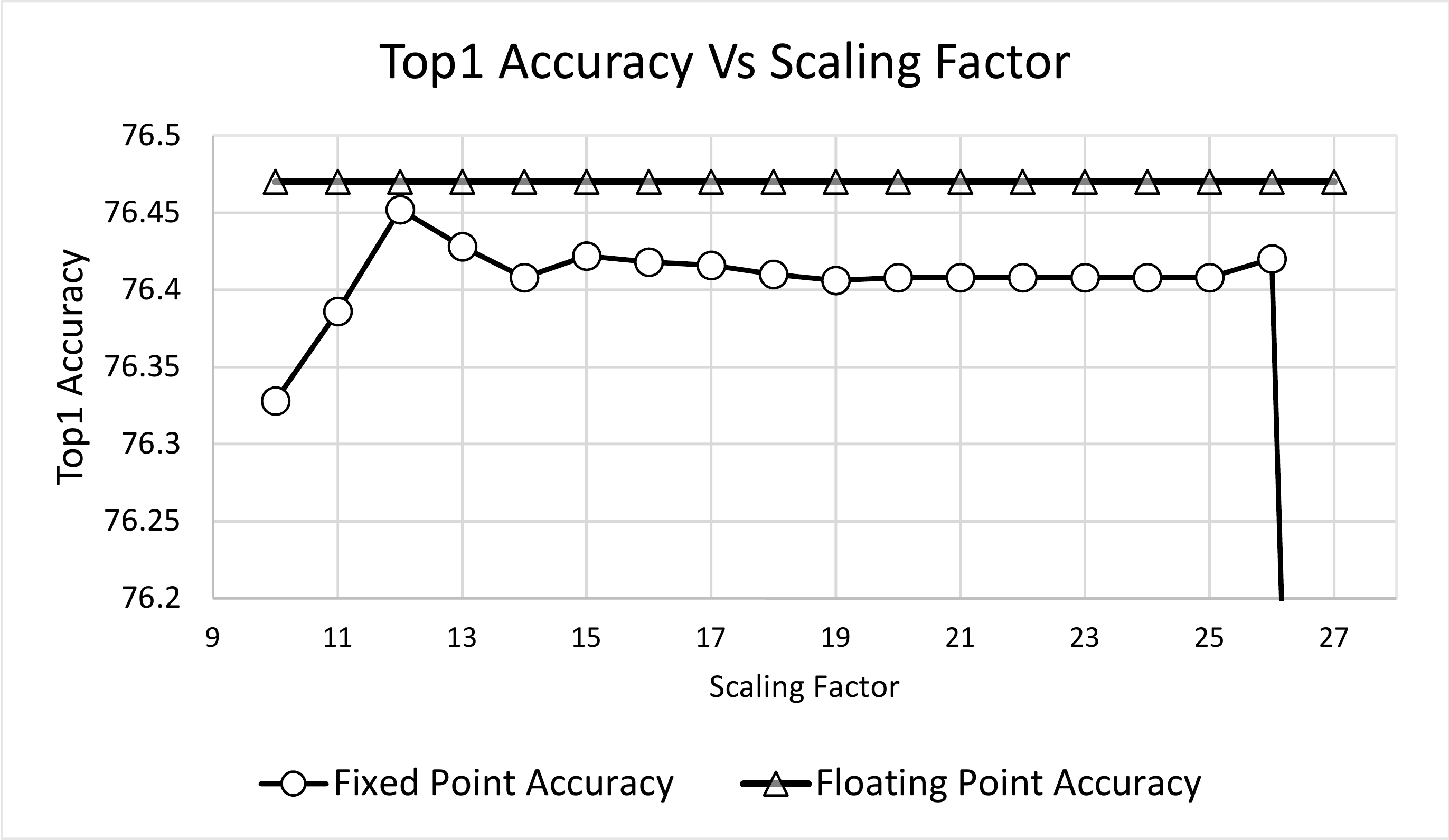}
  \caption{\resnet: Top 1 accuracy vs Scale}
  \label{fig:resnetaccuracy1}
\end{figure}

\begin{figure}
  \includegraphics[width=0.8\linewidth]{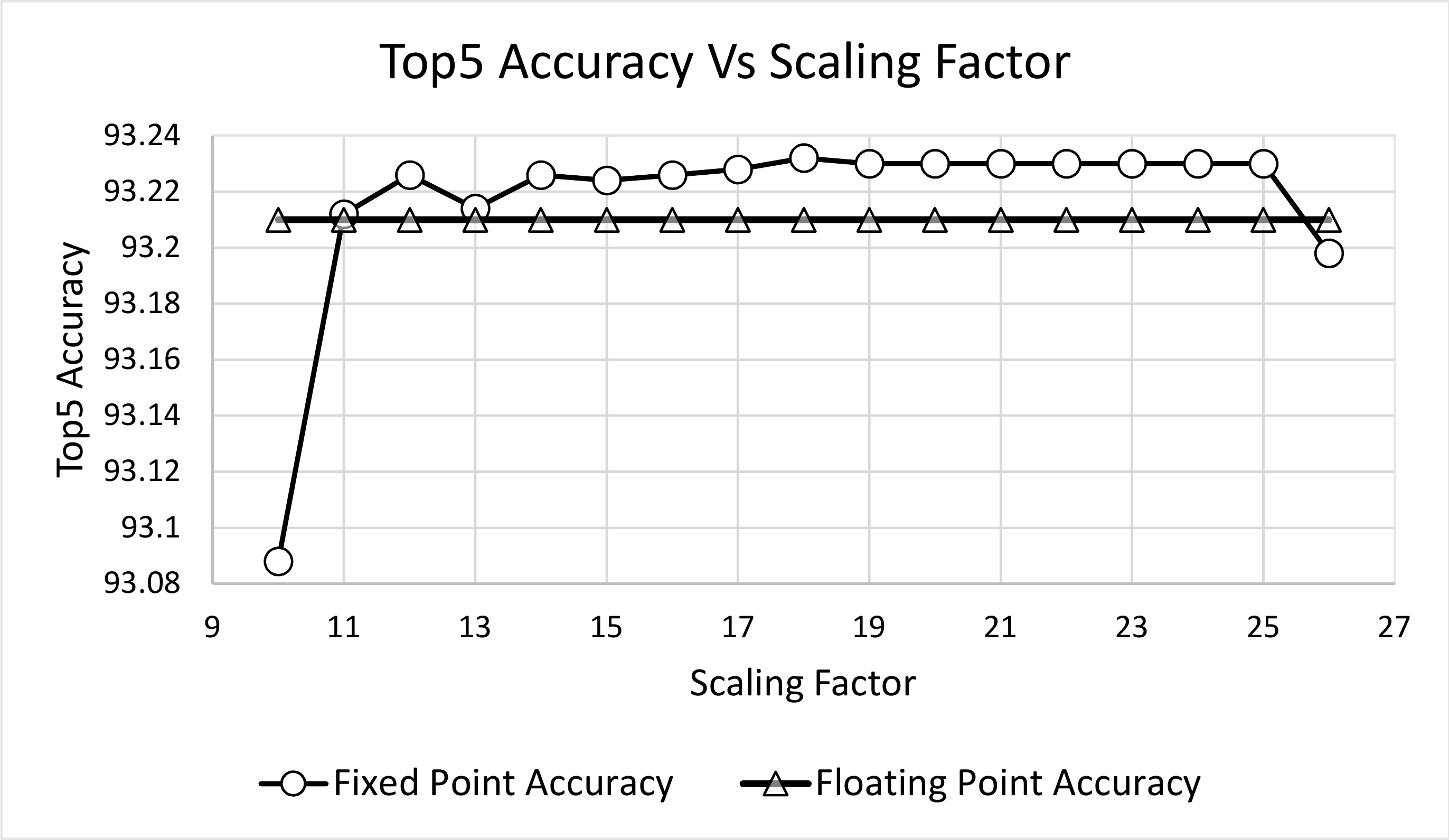}
  \caption{\resnet: Top 5 accuracy vs Scale}
  \label{fig:resnetaccuracy5}
\end{figure}

%\begin{figure}
%  \includegraphics[width=0.8\linewidth]{squeezenetaccuracy1.pdf}
 % \caption{\squeezenet: Top 1 accuracy vs Scale}
  %\label{fig:squeezenetaccuracy1}
%\end{figure}

%\begin{figure}
 % \includegraphics[width=0.8\linewidth]{squeezenetaccuracy5.pdf}
 % \caption{\squeezenet: Top 5 accuracy vs Scale}
  %\label{fig:squeezenetaccuracy5}
%\end{figure}

\begin{figure}
  \includegraphics[width=0.8\linewidth]{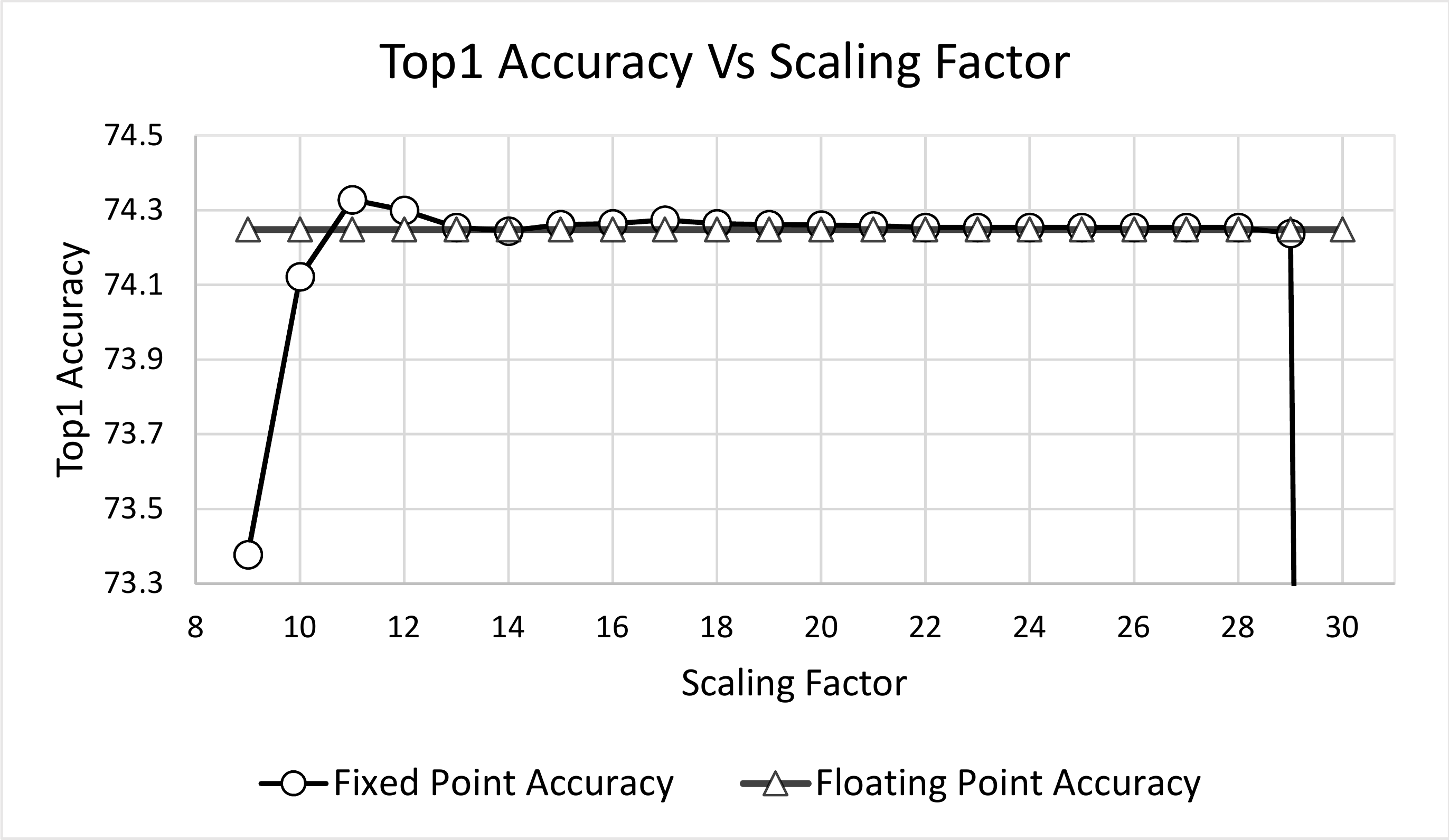}
  \caption{\densenet: Top 1 accuracy vs Scale}
  \label{fig:densenetaccuracy1}
\end{figure}

\begin{figure}
  \includegraphics[width=0.8\linewidth]{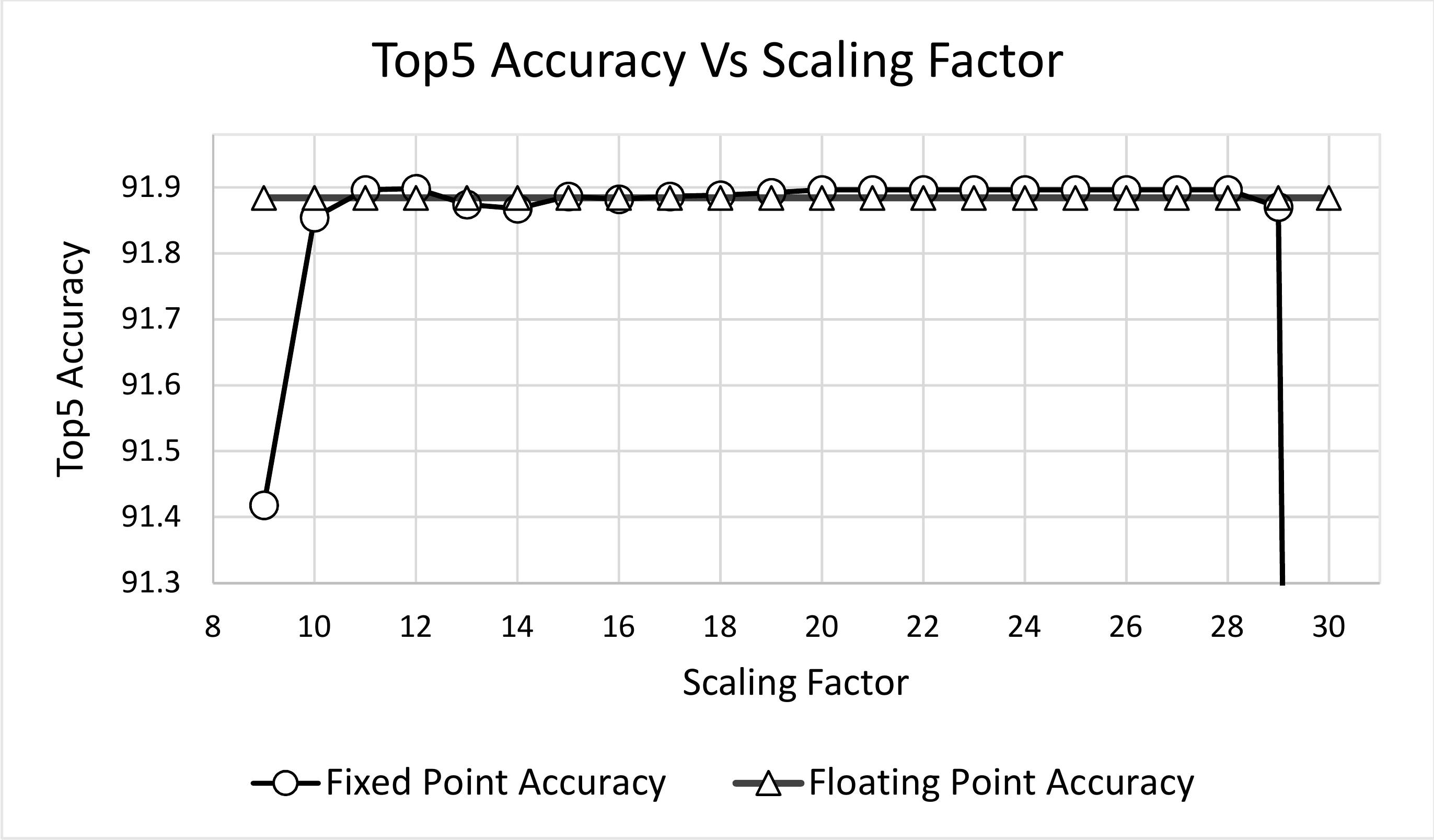}
  \caption{\densenet: Top 5 accuracy vs Scale}
  \label{fig:densenetaccuracy5}
\end{figure}
%\end{comment}
% \begin{figure}
%   \includegraphics[width=0.8\linewidth]{squeezenetaccuracy1.pdf}
%   \caption{\squeezenet: Top 1 accuracy vs Scaling Factor}
%   \label{fig:squeezenetaccuracy1}
% \end{figure}

\subsection{Comparison with 2PC/FHE}
\label{unfaircomp}
See Table~\ref{tab:porthosvspriorcifar10} which validates the well-known fact that 3PC protocols like Porthos are much faster than 2PC/FHE-based approaches.
We omit other 2PC/FHE works (\cite{secureml,ezpc,nhe,nhe2,delphi}, etc.) as the performance comparisons are similar and do not provide additional insights.
\begin{table}
  \centering
%  \resizebox{\columnwidth}{!}{

      \begin{tabular}{|c|c|c|c|c|}
    \hline
    Benchmark & CHET & MiniONN  & Gazelle & Porthos\\
    \hline
	$\squeezenet^*$ (CIFAR) & $1342$ & - & - & $0.05$ \\
	\hline
    MiniONN (CIFAR) & - & $544$  & $12.9$ & $0.36$\\
	\hline
    MiniONN (MNIST) & - & 9.4  & 0.81 & 0.03\\
   \hline 
\end{tabular}
%}
 \caption{Comparison with 2PC  -- All times in seconds. CHET replaces ReLUs in a small \squeezenet\ with square activations.}
\label{tab:porthosvspriorcifar10}
%\tableup
	%\vspace{-0.5cm}
\end{table}
%\mayank{Adding the comparison of Aramis with QuantizedNN here. QuantizedNN uses MP-SDPZ only, so this table serves the purpose of a comparison with both of them.}

\subsection{Proof of malicious security}
\label{app:proof}
For simplicity, consider the case of single malicious party $P_i$. 
Informally, we argue that our technique constrains $P_i$ to follow the instructions of the semi-honest protocol $\pi(\cdot)$ faithfully. 
Or, deviating from faithful execution would result in some honest party to abort. 
%Below, we give a security proof using the standard simulation paradigm (we refer the reader to~\cite{gmw,canetti00} for details on the paradigm).
%
The first $\compute$ invocation of $\fattest^{(\vk_i,\sksign_i)}$ fixes the input of $P_i$ used in the protocol. 
Since every other $\fattest^{(\vk_j,\sksign_j)}$ reliably knows
the verification key $\vk_i$ used by $\fattest^{(\vk_i,\sksign_i)}$,
it checks the signatures on the function description (i.e., $\token^{(i)}_{\pi^*}$)
 as well as the messages of the protocol. The unforgeability of the signature scheme guarantees that $P_i$ cannot forge signatures on incorrectly generated protocol messages. Note that we use this property to ensure that both of the following signatures cannot be forged: (a) signatures under $\vk_i$ on messages generated by $\fattest^{(\vk_i,\sksign_i)}$ and sent to honest $P_j$ (b) signatures under $\vk_j$ on messages sent by $P_j$ being fed into $\fattest^{(\vk_i,\sksign_i)}$. 
Also, $\fattest^{(\vk_i,\sksign_i)}$ provides correct randomness to generate messages of $P_i$ in the semi-honest secure protocol. 
Hence, all messages from $P_i$ to any honest party $P_j$ are generated correctly as directed by $\pi$. 
This argument can be easily extended to multiple colluding corrupt parties.

%Below, we give a security proof using the standard simulation paradigm (we refer the reader to~\cite{gmw,canetti00} for details on the paradigm).

Formally, we give a security proof using the standard simulation paradigm (we refer the reader to~\cite{gmw,canetti00} for details on the paradigm).
%we prove simulation based security, %(Section~\ref{sec:sim-security}),  
That is, the protocol in Figure~\ref{fig:shtomprotocol} securely realizes the ideal \mpc functionality described in Figure~\ref{fig:fideal} against malicious adversaries. 

\begin{theorem}[Restated]
\label{theorem:maliciousmpc}
Let $\pi(\cdot)$ be a semi-honest secure MPC protocol securely realizing $\fmpc^f$. Then, protocol $\protshtom$ described in Figure \ref{fig:shtomprotocol} securely realizes $\fmpc^f$ in the $\fattest^{(\vk_i,\sksign_i)}-$hybrid model (with $i \in [n]$) against malicious adversaries.
\end{theorem}

\newcommand{\simush}{\simu'}

\noindent\begin{proof}[Proof Sketch]  Let $\adv$ be the real world adversary. 
Ideal world adversary $\simu$ that simulates the view of $\adv$ is as follows: 
Let $\simush$ be the ideal world adversary or the semi-honest simulator for $\pi$ (this exists because $\pi$ is semi-honest secure). 
$\simu$ picks $\{(\vk_k,\sksign_k)\}_{k\in[n]}$ and gives $\{\vk_k\}_{k \in [n]}$ to $\adv$. 
We denote a corrupt party by $P_i$ and honest party by $P_j$.
Next, when $\adv$ invokes an instance of $\fattest^{(\vk_i, \sksign_i)}$ on command $\gcommit$ for a corrupted party $P_i$, $\simu$ simulates the correct behavior of $\fattest^{(\vk_i,\sksign_i)}$. 
Also, $\simu$ sends correctly generated tokens $\{\token^{(j)}_{\pi^*}\}$ for all honest parties to $\adv$. 
When $\simu$ receives token from $\adv$ corresponding to a corrupted party $P_i$, it checks it against $\pi^*$ and $\vk_i$. It aborts if verification fails.
When $\adv$ invokes $\fattest^{(\vk_i, \sksign_i)}$ with $x_i$, $\simu$ stores it as input of $P_i$. When $\adv$ commits to inputs of all corrupt parties, $\simu$ sends these to $\fmpc^f$ to learn output $y$. It sends inputs of corrupt parties and outputs $y$ to $\simush$ that generates the view of the adversary in the semi-honest protocol, that contains the randomness for all corrupt parties as well as the transcript of the protocol. Using this, it is easy for $\simu$ to simulate the view of $\adv$ in the rest of the protocol. 
The indistinguishability of the adversary's view in real and ideal executions follows from the semi-honest security of $\pi$. 
\end{proof}

\end{document}